\documentclass{article}

\usepackage{times}
\usepackage{soul}
\usepackage{url}
\usepackage[hidelinks]{hyperref}
\usepackage[utf8]{inputenc}
\usepackage[small]{caption}
\usepackage{graphicx}
\usepackage{amsmath}
\usepackage{amsthm}
\usepackage{booktabs}
\usepackage{algorithm}
\usepackage{algorithmic}
\usepackage{algorithm}
\usepackage{algorithmic}

\newcommand{\removed}[1]{}
\usepackage{times}
\usepackage{soul}
\usepackage{url}
\usepackage[utf8]{inputenc}
\usepackage{graphicx}
\usepackage{amsmath}
\usepackage{amsthm}
\usepackage{booktabs}
\usepackage{algorithm}
\usepackage{algorithmic}
\usepackage{amsmath}
\usepackage{pgfplots}

\usepackage{amsmath,bm}

\usepackage{algorithm}
\usepackage{algorithmic}

\usepackage{times}

\usepackage{url}

\usepackage{amsmath}

\usepackage{amssymb}
\usepackage{mathtools}
\usepackage{amsthm}

\usepackage{algorithmic}

\theoremstyle{plain}

\usepackage[textsize=tiny]{todonotes}
\usepackage{multirow}

\usepackage{ascmac}
\usepackage{float}
\usepackage{perpage}
\MakeSorted{figure}
\MakeSorted{table}

\usepackage{url}
\usepackage{natbib}
\usepackage{chapterbib}

\usepackage{color}
\usepackage{tikz}
\tikzset{%
mynode/.style={circle,minimum width=.5ex, fill=none,draw}, 
myfillnode/.style={circle,minimum width=.5ex, fill=lightgray,draw}, 
}
\usepackage{amssymb}
\usepackage{natbib}

\newcommand{\indep}{\perp \!\!\! \perp}
\usepackage{amsmath}               
\usepackage{lscape}
\usepackage{algorithm}
\usepackage{color}
\usepackage{tikz}
\usepackage{amsmath,amsthm}
\newtheorem{theorem}{Theorem}
\newtheorem{definition}{Definition}
\newtheorem{assumption}{Assumption}

\newtheorem{proposition}{Proposition}
\newtheorem{corollary}{Corollary}

\usepackage{stackengine}
\def\defeq{\mathrel{\ensurestackMath{\stackon[1pt]{=}{\scriptscriptstyle\Delta}}}}

\pgfplotsset{compat=1.18}
\usepackage{comment}
\usepackage{here}
\allowdisplaybreaks[4]
\usepackage{caption}
\usepackage{bbding}
\usepackage{afterpage}
\usepackage{thmtools}
\usepackage{mathrsfs}

\usepackage{wrapfig}

\usepackage{xcolor}
\newcommand{\jin}[1]{\textcolor{blue}{[[#1]]}}
\newcommand{\jina}[1]{\textcolor{blue}{#1}}

\usepackage{soul}

\usepackage[main, final]{neurips_2026}

\newcommand{\dotline}{%
  \noalign{\hbox to \dimexpr 12.2cm+8\tabcolsep+\arrayrulewidth\relax{%
    \leaders\hbox to 2pt{\hss\rule{0.5pt}{0.4pt}\hss}\hfill}}}
\usepackage[utf8]{inputenc} 
\usepackage[T1]{fontenc}    
\usepackage{hyperref}       
\usepackage{url}            
\usepackage{booktabs}       
\usepackage{amsfonts}       
\usepackage{nicefrac}       
\usepackage{array}
\usepackage{microtype}      
\usepackage{xcolor}         
\usepackage{array}
\title{Identification and Bounding of Joint Expectation over Potential Outcomes}

\author{%
Yuta Kawakami and Jin Tian \\
Mohamed bin Zayed University of Artificial Intelligence\\
  \texttt{\{Yuta.Kawakami,Jin.Tian\}@mbzuai.ac.ae} \\
}

\begin{document}

\maketitle

\begin{abstract}
The joint expectation of functions of potential outcomes $(Y_1,Y_0)$ is a fundamental quantity in causal inference, particularly when investigating the variation or heterogeneity of causal effects. This paper provides novel  assumptions for their identification:  comonotonicity and countercomonotonicity. These assumptions offer a unified framework for their identification  across discrete, continuous, and mixed outcome variables. In the absence of these assumptions, we derive sharp bounds for two broad classes of functions, yielding new results for bounding moments of individual causal effects, which are central to measuring effect heterogeneity. We present corresponding estimation methods and illustrate them in simulation studies and real-world datasets.
\end{abstract}

\section{Introduction}

The joint expectation of a function $g(Y_1,Y_0)$  over potential outcomes (POs)  is a fundamental quantity in causal inference, particularly for investigating the variation and heterogeneity of causal effects between an outcome ($Y$) and a binary treatment ($X$), where $Y_x$ denotes PO $Y_{X=x}$. 
The joint expectation over POs is generally defined as
$\mathbb{E}[g(Y_1,Y_0)]\defeq\int g(y_1,y_0)dF_{(Y_1,Y_0)}(y_1,y_0)$,
where $F_{(Y_1,Y_0)}(y_1,y_0)$ represents the joint cumulative distribution function (CDF) of POs. 
The joint expectation over POs plays a central role in characterizing various statistical aspects of causal effects such as moments, CDF, and quantiles, including those defined by the individual causal difference effect (ICDE), i.e., $Y_1-Y_0$, and the individual causal ratio effect (ICRE), i.e., $Y_1/Y_0$.  
Recently, the moments of ICDE have been studied by \citet{Kawakami2025_moments}, 
who established that the moments of ICDE are identifiable under a monotonicity assumption 
and also provided corresponding bounds when the monotonicity is violated.
On the other hand, although 
ICRE has been historically discussed \citep{Rosenbaum1983,Greenland1999,Hernan2024}, its moments—even the mean—have been rarely studied due to the difficulty in identifying and estimating them. 
These forms of counterfactual reasoning are substantially more complex, as they reside in the highest layer—${\cal L}_3$—of Pearl’s causal hierarchy \citep{Bareinboim2022,Pearl2018}.

{The joint expectations over POs may be identified by  identifying the value of unobserved counterfactual outcomes. This can be achieved based on rank preservation assumptions \citep{Heckman1997,Athey2006,Wu2025}, or strict monotonicity plus homogeneity assumptions \citep{Xie2023,Nasr2023}. However, all these assumptions are quite restrictive.}

In this paper, we present new assumptions, comonotonicity and countercomonotonicity, for identifying the joint expectations over POs. These assumptions offer a unified identification framework for joint expectations over POs across discrete, continuous, and mixed outcome variables.
When the outcome variable is continuous, the comonotonicity  reduces to the rank preservation assumption in \citep{Heckman1997}. 
In general, the comonotonicity assumption is weaker than  the rank preservation assumptions \citep{Heckman1997,Wu2025} and the monotonicity assumptions 
\citep{Xie2023,Kawakami2025_moments}.
When $Y$ is binary, the comonotonicity assumption is weaker than the monotonicity assumption in \citep{Pearl1999,Tian2000}.
Table \ref{tab:my_label} summarizes the related identification assumptions  and the corresponding targets of each study.


\begin{table*}[tb]
    \centering
    \caption{Summary of Identification Assumptions and Their Objectives. 
    }
    \label{tab:my_label}
    \begin{tabular}{>{\centering\arraybackslash}p{3.6cm}|>{\centering\arraybackslash}p{2cm}>{\centering\arraybackslash}p{3.35cm}>{\centering\arraybackslash}p{3.25cm}}
    \hline
        Research &  Outcome Type & Assumption &  Objective \\
        \hline\hline
        p. 504 in \citep{Heckman1997} & Continuous & Rank Preservation & Values of Unobserved POs \\
         \dotline
         Assumption 4.6 in \citep{Wu2025} & Mixed & Rank Preservation with $\tilde{\rho}$ & Values of Unobserved POs\\
        \dotline
        Assumption of Theorem 1 in \citep{Xie2023} & Continuous & Strict monotonicity over $f_Y$ & Values of Unobserved POs\\
        \hline
    Definition 14 in \citep{Tian2000} & Binary & Monotonicity & PoC\\
        \dotline
        Assumption 2 in \citep{Kawakami2025_moments} & Continuous & Monotonicity over $f_Y$ & Moments of ICDE \\
        \hline
       Assumptions \ref{comono} and \ref{countercomono} in this paper  & Mixed & Comonotonicity and Countercomonotonicity & Joint Expectations over POs\\
         \hline
    \end{tabular}
\end{table*}

Our assumptions, much like those of monotonicity and rank preservation, may sometimes be implausible for researchers.
Therefore, we next study bounding the joint expectations over POs in the absence of these assumptions.
We obtain sharp bounds for joint expectation of two classes of $g$ functions, and show that the results provide 
tighter bounds for moments of ICDE and, to the best of our knowledge, the first bounds for moments of ICRE. 

{Finally, we present  methods for estimating the joint expectation over POs and their bounds from finite samples, and illustrate them in simulation studies and two real-world datasets.}



{The main contributions of the paper are as follows:}
\begin{itemize}
\item We introduce two new identification assumptions, comonotonicity and countercomonotonicity, for $\mathbb{E}[g(Y_1, Y_0)]$. These assumptions are often weaker than existing alternatives and provide a unified framework applicable to discrete, continuous, and mixed outcome $Y$.
\item We derive novel sharp bounds for $\mathbb{E}[g(Y_1, Y_0)]$ for two broad classes of functions $g$, leading to tighter bounds for moments of ICDE and the first  bounds for moments of ICRE.
\item We develop corresponding estimation procedures for implementation with finite samples.
\end{itemize}

\section{Notation and Background}

We represent each variable with a capital letter $(X)$ and its realized value with a small letter $(x)$.
Let $\mathbb{I}(x)$ be an indicator function that takes $1$ if $x$ is true and $0$ if $x$ is false.
Denote $\Omega_Y$ as the domain of $Y$,
$\mathbb{E}[Y]$ as the expectation of $Y$, and $F_Y(y)=\mathbb{P}(Y\leq y)$ as the cumulative distribution function (CDF)
of
$Y$.
The generalized inverse of the CDF is defined as $F^{-1}_Y(p)=\inf\{y;F_Y(y)\geq p\}$ for $p \in [0,1]$.

\noindent{\bf Comonotonicity and countercomonotonicity.}
Comonotonicity and countercomonotonicity are foundational concepts in the study of dependence structures between random variables \citep{Nelsen2006}.
A pair of random variables $(X,Y)$ is said to be comonotonic if there exists a random variable $U \sim \text{Unif}(0,1)$ and monotonic increasing functions $\alpha$ and $\beta$ such that $X = \alpha(U)$ and $Y = \beta(U)$.
A pair of random variables $(X,Y)$ is said to be countercomonotonic if  there exists a random variable $U \sim \text{Unif}(0,1)$ and 
{monotonic functions $\alpha$ and $\beta$, one decreasing and the other increasing,} 
such that $X=\alpha(U)$ and $Y=\beta(U)$.

\noindent{\bf Causal Effects.}
We use the language of Structural Causal Models (SCM) 
 as our basic semantic and inferential framework \citep{Pearl09}. We denote the PO of $Y$ under intervention $do(X={x})$ by $Y_{{x}}$. 
Researchers often consider the following SCM, ${\cal M}$:
\begin{equation}
Y:=f_Y(X,U_Y), X:=f_X(U_X),
\end{equation}
where $U_Y$ and $U_X$ are latent exogenous variables following a joint distribution $\mathbb{P}(U_Y, U_X)$.
$Y$ is a discrete, continuous, or mixed variable, and $X$ is a binary variable taking values in $\{0, 1\}$. 
The individual causal difference effect (ICDE) is defined as \citep{Holland1986}:
\begin{equation}
\text{\normalfont ICDE}\defeq Y_1-Y_0,
\end{equation}
and the average causal difference effect is defined as $\text{\normalfont ACDE}\defeq\mathbb{E}[Y_1-Y_0]$.
{For outcome variables that take strictly positive values,} the individual causal ratio effect (ICRE) is defined as \citep{Rosenbaum1983,Greenland1999,Hernan2024}:
\begin{equation}
\text{\normalfont ICRE}\defeq {Y_1}/{Y_0}.
\end{equation} 
The average causal ratio effect (ACRE) is given by  
$\text{\normalfont ACRE}\defeq \mathbb{E}\left[{Y_1}/{Y_0}\right]$.
The identification of ACRE is more difficult than the identification of ACDE because ACRE is not the causal risk ratio \citep{Hernan2006}, i.e., 
$\mathbb{E}\left[{Y_1}/{Y_0}\right] \ne {\mathbb{E}[Y_1]}/{\mathbb{E}[Y_0]}$. 

\section{Joint Expectations over POs}

First, we show the formal definition of 
the joint expectation of a function $g$ over POs. 
\begin{definition}
The joint CDF of $(Y_1,Y_0)$ is defined as 
$F_{(Y_1,Y_0)}(y_1,y_0)\defeq\mathbb{P}(Y_1\leq y_1,Y_0 \leq y_0)$.
\end{definition}
The target of this paper is the joint expectation of a function $g$ evaluated over the POs $(Y_1,Y_0)$.
\begin{definition}
The joint expectation of a measurable function $g$ over POs $(Y_1,Y_0)$ is defined as 
$\mathbb{E}[g(Y_1,Y_0)]\defeq\int g(y_1,y_0)dF_{(Y_1,Y_0)}(y_1,y_0)$.
\end{definition}
\noindent 
When the function $g(y_1, y_0)$ takes the separable form $g_1(y_1) - g_0(y_0)$, as in the case of the ACDE, computing its expectation does not require the joint distribution of $(Y_1, Y_0)$. 
Specifically, the expectation simplifies as
$\mathbb{E}[g(Y_1, Y_0)] = \int g_1(y_1) dF_{Y_1}(y_1) - \int g_0(y_0) dF_{Y_0}(y_0)$,
which depends only on the marginals $Y_1$ and $Y_0$.

\subsection{Why do  joint expectations over POs matter?}

Joint expectations over POs play a central role in capturing various statistical aspects of the variation or heterogeneity in causal effects for both ICDE and ICRE.
Firstly, we can represent the CDF of ICDE and ICRE in the form of a joint expectation over POs as follows:
\begin{align}
&\mathbb{P}(Y_1-Y_0<y)=\int\mathbb{I}(y_1-y_0<y)dF_{(Y_1,Y_0)}(y_1,y_0),\label{eq8}\\
&\mathbb{P}(Y_1/Y_0<y)=\int\mathbb{I}(y_1/y_0<y)dF_{(Y_1,Y_0)}(y_1,y_0).\label{eq10}
\end{align}
Additionally, the moments of ICDE and ICRE are special cases of joint expectations over POs with $g(Y_1, Y_0) = (Y_1-Y_0)^m$ and $g(Y_1, Y_0) = (Y_1/Y_0)^m$. So are their central moments with $g(Y_1, Y_0) = \{(Y_1-Y_0)-(\mathbb{E}[Y_1]-\mathbb{E}[Y_0])\}^m$, and $g(Y_1, Y_0) = \{(Y_1/Y_0)-(\mathbb{E}[Y_1/Y_0])\}^m$. 
These moments encompass statistical measures such as mean, variance,
skewness, and kurtosis, and have recently been studied by \citet{Kawakami2025_moments} in the context of ICDE. They go beyond the average causal effect (which represents the first moment) and characterize the shape of the distribution of causal effects. 
The shape of the distribution of causal effects uncovers causal effect heterogeneity.
In the context of ICDE, existing studies on effect heterogeneity focus on the conditional average treatment effect (CATE), defined as $\mathbb{E}[Y_1-Y_0|W]$, where $W$ denotes observed covariates. 
However, CATE captures heterogeneity only at the level of subpopulations defined by $W$, while 
these joint expectations over POs capture heterogeneity at the individual level, providing insights beyond subgroup averages.

Furthermore, the probability of necessity and sufficiency (PNS) for a continuous outcome \citep{Kawakami2024} is encompassed within the joint expectation over POs, i.e., $\mathbb{P}(Y_0< y\leq Y_1)=\int \mathbb{I}(y_0< y\leq y_1)dF_{(Y_1,Y_0)}(y_1,y_0)$.
PNS reveals another aspect of the causal relationship.
As another example, 
the causal parameter $\mathbb{P}(Y_1> Y_0)$ studied by \citet{Fay2018}, which is the probability that a randomly selected individual will have a larger response under action than under control, 
is also included within the joint expectation over POs, i.e., $\mathbb{P}(Y_1> Y_0)=\int \mathbb{I}(y_1> y_0)dF_{(Y_1,Y_0)}(y_1,y_0)$.
This causal parameter has received considerable attention in the field of causal inference \citep{Gadbury2004,Fan2010,Huang2016,Lu2019_bound,Gabriel2024}.  

\section{Identification of Joint Expectations over POs}


We begin by presenting the Fr\'{e}chet–Hoeffding bounds on the joint CDF of POs.
\begin{restatable}{lemma}{Lemmaone}
\label{lemma1}
Given $F_{Y_0}$ and $F_{Y_1}$, 
the joint CDF of POs $F_{(Y_1,Y_0)}(y_1,y_0)$ is bounded by $F^L_{(Y_1,Y_0)}(y_1,y_0)\leq F_{(Y_1,Y_0)}(y_1,y_0)\leq F^U_{(Y_1,Y_0)}(y_1,y_0)$, where 
\begin{align}
F^U_{(Y_1,Y_0)}(y_1,y_0)=\min\{F_{Y_0}(y_0),F_{Y_1}(y_1)\}, F^L_{(Y_1,Y_0)}(y_1,y_0)=\max\{F_{Y_0}(y_0)+F_{Y_1}(y_1)-1,0\}.\label{eq-lower}
\end{align}
The upper bound is attained when $(Y_1, Y_0)$ are comonotonic, while the lower bound is attained when $(Y_1, Y_0)$ are countercomonotonic. {These bounds are sharp given  $F_{Y_0}$ and $F_{Y_1}$.}
\end{restatable}
\noindent 
{We note that $F^U_{(Y_1,Y_0)}$ and $F^L_{(Y_1,Y_0)}$ defined by Eq.~\eqref{eq-lower} are valid joint CDFs of $(y_1, y_0)$ \citep{Nelsen2006}. In addition, the marginal distributions of $F^U_{(Y_1,Y_0)}$ and $F^L_{(Y_1,Y_0)}$ over $Y_0$ are both equal to $F^U_{Y_0}=F^L_{Y_0}=F_{Y_0}$ \citep{Nelsen2006}.}

We then introduce new identification assumptions for the joint expectation over POs, corresponding to comonotonicity and countercomonotonicity.
\begin{assumption}[Comonotonicity]
\label{comono}
Potential outcomes $(Y_1,Y_0)$ are comonotonic.
\end{assumption}
\begin{assumption}[Countercomonotonicity]
\label{countercomono}
Potential outcomes $(Y_1,Y_0)$ are countercomonotonic.
\end{assumption}
\noindent 
In terms of $(Y_1, Y_0)$, comonotonicity means that whenever individual $A$ would have a higher outcome than $B$ under $do(X=0)$, $A$ would not have a lower outcome than $B$ under $do(X=1)$, almost surely.
In practice, comonotonicity is plausible when the interventions differ only in intensity, dosage, or delivery mode—such as alternative dosing schedules of the same drug, short versus extended program duration, or online versus in-person delivery of the same service—so that individuals who benefit more under one condition also benefit more under the other. Comonotonicity is also reasonable when treatment effects are additive or proportional, or when both POs are increasing functions of a shared latent trait such as baseline severity 
or responsiveness. {We refer to Appendix~\ref{appE} for real-world applications that concretely illustrate the practical interpretation of comonotonicity.}

{In contrast, countercomonotonicity means that whenever individual $A$ would have a higher outcome than $B$ under $do(X=0)$, $A$ would not have a higher outcome than $B$ under $do(X=1)$, and vice versa.
Countercomonotonicity is plausible when the two interventions benefit different types of individuals in systematically opposing ways, so that those who would perform better under one intervention tend to perform worse under the other. This can arise in remedial education programs that disproportionately benefit low-ability students, in targeted subsidies that primarily support the least productive firms, or in assistive technologies that compensate for physical or cognitive limitations.}

 {We note that the comonotonicity and  countercomonotonicity assumptions are applicable to any type of outcome variables—discrete, continuous, or mixed.}
 


We {define}
\begin{align}
&\mathbb{E}^{U}[g(Y_1,Y_0)]\defeq\int g(y_1,y_0)dF^U_{(Y_1,Y_0)}(y_1,y_0), \label{eq-eug}\\
&\mathbb{E}^{L}[g(Y_1,Y_0)]\defeq\int g(y_1,y_0)dF^L_{(Y_1,Y_0)}(y_1,y_0).\label{eq-eug-l}
\end{align}
Then, the joint expectation over POs is identifiable as below:
\begin{restatable}{theorem}{Theoremone}
\label{theorem2}
Under SCM ${\cal M}$, given $F_{Y_0}$ and $F_{Y_1}$, 
we have the following two identification statements:\vspace{0.1cm}

\noindent (A). Under Assumption \ref{comono},
$\mathbb{E}[g(Y_1,Y_0)]=\mathbb{E}^{U}[g(Y_1,Y_0)]$.

\noindent (B). Under Assumption \ref{countercomono},
$\mathbb{E}[g(Y_1,Y_0)]=\mathbb{E}^{L}[g(Y_1,Y_0)]$.
\end{restatable}

\subsection{Comparison with previous identification assumptions}

We compare  comonotonicity  (Assumption \ref{comono}) with previous identification assumptions in the literature.
We note that countercomonotonicity has rarely been studied in prior work.


\noindent{\bf Strict monotonicity.}
 Strict monotonicity is used for the identification of values of unobserved POs \citep{Lu2020,Xie2023,Nasr2023}.
\begin{assumption}[Strict Monotonicity]
\label{STR}
{The functions $f_Y(1,U_Y)$ and $f_Y(0,U_Y)$ are smooth and strictly monotone in $U_Y$, with the same direction of monotonicity.}
\end{assumption}
\begin{restatable}{proposition}{Propositionone}
\label{propstr}
Assumption \ref{STR} implies Assumption~\ref{comono}; {however, the converse implication does not hold.}
\end{restatable}
\noindent The strict monotonicity assumption is stronger than comonotonicity and is violated if the outcome variable is discrete or mixed.



\noindent{\bf Monotonicity for binary outcome.}
For the binary outcome $Y$, the following monotonicity assumption is often utilized in the probabilities of causation (PoC) literature \citep{Pearl1999,Tian2000}:
\begin{assumption}[Monotonicity for Binary Outcome]
\label{MONO}
$\mathbb{P}(Y_0=1,Y_1=0)=0$ holds.
\end{assumption}
\begin{restatable}{proposition}{Propositiontwo}
\label{probin}
{When $Y$ is binary, Assumption \ref{comono} is equivalent to the condition that either $\mathbb{P}(Y_0=1,Y_1=0)=0$ or $\mathbb{P}(Y_0=0,Y_1=1)=0$ holds, which is weaker than Assumption \ref{MONO}.}
\end{restatable}
\noindent 

\noindent{\bf Monotonicity over $f_Y$.}
Monotonicity over $f_Y$ is used to identify PoC and  moments of ICDE \citep{Kawakami2024,Kawakami2025_moments}.
\begin{assumption}[Monotonicity over $f_Y$]
\label{MONO2}
{The function $f_Y(x,U_Y)$ is either (i) monotonic increasing on $U_Y$ for all $x \in \{0,1\}$ almost surely w.r.t. $\mathbb{P}_{U_Y}$, or (ii) monotonic decreasing on $U_Y$ for all $x \in \{0,1\}$ almost surely w.r.t. $\mathbb{P}_{U_Y}$.} 
\end{assumption}
\begin{restatable}{proposition}{Propositionthree}\label{propmono}
{Assumption \ref{MONO2} implies Assumption \ref{comono}; however, the converse need not hold for a given structural representation $Y=f_Y(X, U_Y)$. Nevertheless, if $(Y_1, Y_0)$ are comonotonic, then there exists a latent variable $U'_Y$ and a structural function $f'_Y(X, U'_Y)$ satisfying Assumption \ref{MONO2} such that $Y_x = f'_Y(x, U'_Y)$ for $x\in \{0, 1\}$.}
\end{restatable}
\noindent Therefore, Assumption \ref{comono} is weaker than Assumption \ref{MONO2}.
\citet{Kawakami2024} also provide an equivalent assumption (monotonicity over $Y_x$) to Assumption \ref{MONO2} under their Assumption 3.6. 

\noindent{\bf Rank preservation assumptions.}
The rank preservation assumption is often used for the identification of the values of the unobserved POs \citep{Heckman1997}.
\begin{assumption}[Rank preservation assumption]
\label{POD} 
For continuous CDFs $F_{Y_1}$ and $F_{Y_0}$,
$F_{Y_1}(y_1)=F_{Y_0}(y_0)$ holds  
for almost every subject whose POs are $(Y_1,Y_0)=(y_1,y_0)$.
\end{assumption}
\noindent {The rank preservation assumption is closely related to  comonotonicity. In fact, they are equivalent when the outcome variable is continuous. The causal inference literature has often employed the rank preservation assumption, but has rarely explicitly considered comonotonicity.}
\begin{restatable}{proposition}{Propositionfour}
\label{prop-equi}
For continuous outcome $Y$,
Assumption~\ref{comono} is equivalent to Assumption~\ref{POD}.
\end{restatable}
\noindent {The rank preservation assumption is only applicable to a continuous outcome. However, the comonotonicity assumption is applicable to any type of variables - discrete, continuous, or mixed. }
\citet{Wu2025} showed a weaker rank preservation assumption that is applicable to any type of variables using Kendall's rank correlation coefficient \citep{Kendall1938}:
\begin{definition}[Modified Kendall Rank Correlation Coefficient]
Let $(x_1,y_1),\dots, (x_n,y_n)$ be the observations of two random variables $(X,Y)$; the modified Kendall rank correlation coefficient is defined as
$\tilde{\rho}_n(X,Y)=\sum_{1\leq i<j\leq n}\frac{\text{\normalfont sign}\{(x_i-x_j)(y_i-y_j)\}}{\sqrt{n(n-1)/2-T_x}\sqrt{n(n-1)/2-T_y}}$,
where $T_x$ is the number of tied pairs in $\{x_1,\dots,x_n\}$, and $T_y$ is the number of tied pairs in $\{y_1,\dots,y_n\}$.
\end{definition}
\begin{assumption}[Rank Preservation on Kendall's $\tilde{\rho}$]
\label{Kendall}
$\tilde{\rho}_n(Y_1,Y_0)=1$ holds for all integers $n \geq 2$. 
\end{assumption}
\noindent The condition $\tilde{\rho}(Y_1, Y_0) = 1$ indicates that $Y_1$ and $Y_0$ are perfectly positively correlated.
Rank preservation with Kendall's $\tilde{\rho}$ is weaker than rank preservation.
Rank preservation with Kendall's $\tilde{\rho}$ accommodates both discrete and continuous outcomes.
Both identification assumptions lead to the same identified results:
Under SCM ${\cal M}$ and Assumptions 
\ref{POD} (or \ref{Kendall}),
given continuous CDF $F_{Y_1}$ and $F_{Y_0}$,
for any subjects whose POs are $(Y_1,Y_0)=(y_1,y_0)$,
$y_1$ is identifiable by 
\begin{equation}
\label{eq16}
y_1=F_{Y_1}^{-1}(F_{Y_0}(y_0)).
\end{equation}
Therefore, 
the joint expectations over POs are identified by 
\begin{equation}\label{eq-rankeq}
\mathbb{E}[g(Y_1,Y_0)]=\mathbb{E}[g(F_{Y_1}^{-1}(F_{Y_0}(Y_0)),Y_0)].
\end{equation}
\noindent 
We note that the right-hand side of Eq.~\eqref{eq-rankeq} is equivalent to the right-hand side of Eq.~\eqref{eq-eug} 
under Assumption 
\ref{POD} (or \ref{Kendall}). 
\noindent We have the following relationship between Assumption \ref{Kendall} and Assumption \ref{comono}:
\begin{restatable}{proposition}{Propositionfive}
\label{prop2}
Assumption \ref{Kendall} implies Assumption \ref{comono}; however, the converse implication does not hold.
\end{restatable}
\noindent 
Hence, Assumption \ref{comono} is weaker than Assumption \ref{Kendall}. 
We note that the values of the unobserved POs are \emph{not} identified in general under our comonotonicity assumption, and Eq.~\eqref{eq-rankeq} 
does not hold in general under Assumption \ref{comono}.

In summary, by Propositions~\ref{propstr}-\ref{prop2}, \emph{our comonotonicity assumption (Assumption \ref{comono}) offers a weaker identifying assumption for joint expectations over POs than  existing assumptions in the literature.} Moreover, \emph{it is applicable to any type of outcome variables - discrete, continuous, or mixed}.

Furthermore, we establish the following new identification theorem for unobserved POs under the countercomonotonicity assumption:
\begin{restatable}{theorem}{Theoremtwo}
\label{theo-counter}
Under SCM ${\cal M}$ and Assumption
\ref{countercomono}, given $F_{Y_0}$ and $F_{Y_1}$, 
when $Y$ is continuous,
for any subjects whose POs are $(Y_1,Y_0)=(y_1,y_0)$,
$y_1$ is identifiable by 
\begin{equation}
\label{eq16con}
y_1=F^{-1}_{Y_1}(1-F_{Y_0}(y_0)).
\end{equation}
\end{restatable}

\section{Bounding Joint Expectations over POs}

Like monotonicity and rank preservation assumptions, our assumptions may sometimes be implausible to researchers.
Therefore, we next study bounding the joint expectations over POs in the absence of these assumptions.
We present two situations in which 
$\mathbb{E}^{U}[g(Y_1,Y_0)]$ and $\mathbb{E}^{L}[g(Y_1,Y_0)]$ provide sharp bounds for the joint expectation over POs $\mathbb{E}[g(Y_1,Y_0)]$. 
Specifically, we consider the following two classes of $g$ functions:
\begin{assumption}[Supermodularity of $g$]
\label{supermod}
The function $g(y_1, y_0)$ is bounded and supermodular, i.e., the inequality $g(y_1',y_0')-g(y_1,y_0')\geq g(y_1',y_0)-g(y_1,y_0)$ holds for every $y_0, y_1, {y'_0, y'_1} \in \Omega_Y$ such that $y_1\leq  y_1'$ and $y_0\leq  y_0'$.
\end{assumption}
\noindent If $g$ is twice differentiable, supermodularity means  $\frac{\partial^2 g(y_1,y_0)}{\partial{y_1}\partial{y_0}}\geq 0$ for any $y_1, y_0 \in \Omega_Y$.
\begin{assumption}[Submodularity of $g$]
\label{submod}
The function $g(y_1, y_0)$ is bounded and submodular, i.e., the inequality $g(y_1',y_0')-g(y_1,y_0')\leq g(y_1',y_0)-g(y_1,y_0)$ holds for every $y_0, y_1, {y'_0, y'_1} \in \Omega_Y$ such that $y_1\leq  y_1'$ and $y_0\leq  y_0'$.
\end{assumption}
\noindent If $g$ is twice differentiable, submodularity means $\frac{\partial^2 g(y_1,y_0)}{\partial{y_1}\partial{y_0}}\leq 0$ for any $y_1, y_0 \in \Omega_Y$.
Then, we have the following bounding theorem:
\begin{restatable}{theorem}{Theoremthree}
\label{theorem1}
Under SCM ${\cal M}$,
{given $F_{Y_0}$ and $F_{Y_1}$, }
we have the following two statements:\vspace{0.2cm}

\noindent \text{(A). } Under Assumption \ref{supermod}, $\mathbb{E}[g(Y_1,Y_0)]$ is bounded by $\mathbb{E}^L[g(Y_1,Y_0)] \leq \mathbb{E}[g(Y_1,Y_0)] \leq \mathbb{E}^U[g(Y_1,Y_0)]$.\vspace{0.2cm}

\noindent \text{(B). } Under Assumption \ref{submod}, $\mathbb{E}[g(Y_1,Y_0)]$ is bounded by $\mathbb{E}^U[g(Y_1,Y_0)] \leq \mathbb{E}[g(Y_1,Y_0)] \leq \mathbb{E}^L[g(Y_1,Y_0)]$.\vspace{0.2cm}

\noindent These bounds are sharp given the marginals $F_{Y_0}$ and $F_{Y_1}$.
\end{restatable}
\noindent Theorem \ref{theorem1} does not provide fully general bounds for an arbitrary function $g$. 
Still, it leads to new bounding results for the moments and central moments of ICDE and ICRE, 
as discussed in the sequel.

\subsection{Bounds for the moments of ICDE}
First, both the functions $g(y_1,y_0)=(y_1-y_0)^m$ and $g(y_1,y_0)=\{(y_1-y_0)-(\mathbb{E}[Y_1]-\mathbb{E}[Y_0])\}^m$ are submodular when $m$ is even.
\begin{corollary}
\label{cor-1}
Under SCM ${\cal M}$,
{if $Y_1$ and $Y_0$ are bounded,}
the even-order moments of the ICDE are bounded by $\mathbb{E}^U[(Y_1 - Y_0)^m]\leq \mathbb{E}[(Y_1 - Y_0)^m]\leq \mathbb{E}^L[(Y_1 - Y_0)^m]$, and the bounds are sharp.
\end{corollary}
\begin{corollary}
Under SCM ${\cal M}$,
if $Y_1$ and $Y_0$ are bounded, 
the even-order central moments of the ICDE are bounded by $\mathbb{E}^U[\{(Y_1-Y_0)-(\mathbb{E}[Y_1]-\mathbb{E}[Y_0])\}^m]\leq \mathbb{E}[\{(Y_1-Y_0)-(\mathbb{E}[Y_1]-\mathbb{E}[Y_0])\}^m]\leq \mathbb{E}^L[\{(Y_1-Y_0)-(\mathbb{E}[Y_1]-\mathbb{E}[Y_0])\}^m]$, and the bounds are sharp.
\end{corollary}
\noindent 
The sharp upper bound above coincides with that provided by \citet{Kawakami2025_moments}. 
Our sharp lower bound is tighter than theirs, as their lower bound is not sharp.

On the other hand, when $m$ is odd, the $m$-th moment of ICDE can be given by $\mathbb{E}[(Y_1-Y_0)^m] = \sum_{l=0}^m \binom{m}{l}(-1)^l\mathbb{E}[Y_1^{m-l}Y_0^l]$.
If $Y_1$ and $Y_0$ take non-negative values, then for any $k, l$, each $\mathbb{E}[Y_1^kY_0^l]$ is bounded as $\mathbb{E}^L[Y_1^kY_0^l] \leq \mathbb{E}[Y_1^kY_0^l] \leq \mathbb{E}^U[Y_1^kY_0^l]$ since the function $g(y_1,y_0)=y_1^ky_0^l$ is supermodular as $\frac{\partial^2 (y_1^ky_0^l)}{\partial{y_1}\partial{y_0}}=kly_1^{k-1}y_0^{l-1}\geq 0$.
\begin{corollary}\label{cor-3}
Under SCM ${\cal M}$,
{if $Y_1$ and $Y_0$ are bounded and take non-negative values,}
the odd-order moments of the ICDE are bounded by
\begin{equation}
\begin{aligned}
&\sum_{l=0}^m \binom{m}{l}(-1)^l\big\{\mathbb{I}((-1)^l=-1)\mathbb{E}^U[Y_1^{m-l}Y_0^l]+\mathbb{I}((-1)^l=1)\mathbb{E}^L[Y_1^{m-l}Y_0^l]\big\}\leq \mathbb{E}[(Y_1 - Y_0)^m]\\
&\leq \sum_{l=0}^m \binom{m}{l}(-1)^l\big\{\mathbb{I}((-1)^l=-1)\mathbb{E}^L[Y_1^{m-l}Y_0^l]+\mathbb{I}((-1)^l=1)\mathbb{E}^U[Y_1^{m-l}Y_0^l]\big\}.
\end{aligned}
\end{equation}
\end{corollary}
\noindent The bounds for the odd-order central moments of ICDE are presented in Appendix \ref{appB}. These bounds are not sharp.

\subsection{Bounds for the moments of ICRE}
Theorem \ref{theorem1} provides bounds for the moments and central moments of ICRE, {where the outcome variable is assumed to take strictly positive values (otherwise the ratio effect is not meaningful).} 
{To our knowledge, these bounds are novel and represent a new contribution to the literature.}

First, $g(y_1,y_0) = (y_1/y_0)^m$ is submodular for any $m$ when 
{$Y_1/ Y_0$ takes positive values}. We have
\begin{corollary}\label{cor-4}
Under SCM ${\cal M}$,
{if $Y_1/Y_0$ is bounded and takes positive values,}  
the moments of the ICRE are bounded by $\mathbb{E}^U[(Y_1/Y_0)^m]\leq \mathbb{E}[(Y_1/Y_0)^m]\leq \mathbb{E}^L[(Y_1/Y_0)^m]$, and the bounds are sharp.
\end{corollary}
\noindent The  central moments of ICRE are  bounded as follows: 
\begin{corollary}
\label{cor-5}
Under SCM ${\cal M}$,
{if $Y_1/Y_0$ is bounded and takes positive values,}  
the central moments of the ICRE are bounded by
{
\begin{equation}
\begin{aligned}
&\sum_{l=0}^m \binom{m}{l}\big(\mathbb{I}((-1)^l=1)\left( \mathbb{E}^U[Y_1/Y_0] \right)^l\mathbb{E}^U[(Y_1/Y_0)^{m-l}]-\mathbb{I}((-1)^l=-1)\\
&\hspace{8cm}\times\left( \mathbb{E}^L[Y_1/Y_0] \right)^l\mathbb{E}^L[(Y_1/Y_0)^{m-l}]\big)\\
&\leq\mathbb{E}[\{(Y_1/Y_0)-(\mathbb{E}[Y_1/Y_0])\}^m]\\
&\leq\sum_{l=0}^m \binom{m}{l}\big(\mathbb{I}((-1)^l=1)\left( \mathbb{E}^L[Y_1/Y_0] \right)^l\mathbb{E}^L[(Y_1/Y_0)^{m-l}]-\mathbb{I}((-1)^l =-1)\\
&\hspace{8cm}\times\left(\mathbb{E}^U[Y_1/Y_0] \right)^l\mathbb{E}^U[(Y_1/Y_0)^{m-l}]\big).
\end{aligned}    
\end{equation}
}
\end{corollary}
\noindent 
The above bounds are not sharp.

Finally, given the bounds for the 2nd, 3rd, and 4th central moments of ICRE,  the skewness and kurtosis of ICRE can be bounded as shown in Appendix \ref{appB}. 

\textbf{Comparison with existing bounding results.} 
{There is a substantial literature on deriving closed-form  bounds for causal quantities under various assumptions (e.g., \citep{Manski1997,Dawid2016,Sachs2023,Zhang2025}). Relative to \citep{Kawakami2025_moments}, which represents the current state of the art for moments of ICDE, we obtain sharper bounds for these quantities.  To the best of our knowledge, our work provides the first set of bounds for moments of ICRE, which have not been directly addressed in the prior literature. Related approaches compute bounds via numerical optimization, including methods based on  optimal transport \citep{Gao2025}, linear programming \citep{Hashimoto2026}, and polynomial programming in general causal graphs  \citep{Zhang2022_Partial,zaffalon2024efficient,Duarte2024}.
}



\section{Estimating Joint Expectations over POs}

In this section, we illustrate estimation methods for $\mathbb{E}^U[g(Y_1, Y_0)]$ and $\mathbb{E}^L[g(Y_1, Y_0)]$,  which are used to identify and bound the joint expectation over POs. We leverage inverse transform sampling (ITS) \citep{Mcneil2015}, 
a method for generating samples from a target distribution by applying the inverse of its CDF to samples drawn from the uniform distribution on $[0,1]$. 

From the Radon–Nikodym theorem \citep{Billingsley2017}, Eqs.~\eqref{eq-eug} and \eqref{eq-eug-l} can be rewritten as follows:
{
\begin{align}
\label{eq-23}
&\mathbb{E}^U[g(Y_1,Y_0)]=\int \int g(y_1,y_0)dF^U_{Y_1|Y_0}(y_1,y_0)dF_{Y_0}(y_0),\\
\label{eq-24}
&\mathbb{E}^L[g(Y_1,Y_0)]=\int \int g(y_1,y_0)dF^L_{Y_1|Y_0}(y_1,y_0)dF_{Y_0}(y_0),  
\end{align}}
{where $F^U_{Y_1|Y_0}(y_1,y_0)$ and $F^L_{Y_1|Y_0}(y_1,y_0)$ denote conditional CDFs, and we have used $F^U_{Y_0}=F^L_{Y_0}=F_{Y_0}$.  These  hold for discrete, continuous, and mixed-type outcomes. }

First, we derive the explicit expression of 
the conditional CDFs $F^U_{Y_1|Y_0}(y_1,y_0)$ and $F^L_{Y_1|Y_0}(y_1,y_0)$ from the joint CDFs given by Eq.~\eqref{eq-lower}.
We use $F'$ to denote CDFs without the point $y_0$ as follows: 
\begin{align}
F'_{Y_0}(y_0)\defeq\mathbb{P}(Y_0<y_0).
\end{align}
We obtain the following lemma:
\begin{restatable}{lemma}{Lemmatwo}
\label{lemma2}
The conditional CDFs of $F^U_{Y_1|Y_0}(y_1,y_0)$ and $F^L_{Y_1|Y_0}(y_1,y_0)$ are given as follows:\vspace{0.2cm}\\
(A). When $F_{Y_0}(y_0)\ne F'_{Y_0}(y_0)$, which means $y_0$ is a mass point of $F_{Y_0}$, we have
{
\begin{align}
\footnotesize
\label{eq26}
&F^U_{Y_1|Y_0}(y_1,y_0)=\max\left\{\frac{\min\left\{F_{Y_1}(y_1),F_{Y_0}(y_0)\right\}-F'_{Y_0}(y_0)}{F_{Y_0}(y_0)-F'_{Y_0}(y_0)},0\right\},\\
\label{eq27}
&F^L_{Y_1|Y_0}(y_1,y_0)=1-\max\left\{\frac{\min\left\{1-F_{Y_1}(y_1),F_{Y_0}(y_0)\right\}-F'_{Y_0}(y_0)}{F_{Y_0}(y_0)-F'_{Y_0}(y_0)},0\right\}.
\end{align}}
(B). When $F_{Y_0}(y_0)= F'_{Y_0}(y_0)$, which means $y_0$ is not a mass point of $F_{Y_0}$, we have
\begin{align}
&F^U_{Y_1|Y_0}(y_1,y_0)=\mathbb{I}(y_1\geq F_{Y_1}^{-1}(F_{Y_0}(y_0))), F^L_{Y_1|Y_0}(y_1,y_0)=\mathbb{I}(y_1\geq F_{Y_1}^{-1}(1-F_{Y_0}(y_0))).
\end{align}
\end{restatable}


\noindent To apply ITS, we compute the inverse functions of the above conditional CDFs and obtain the following theorem:
\begin{restatable}{theorem}{Theoremfour}
\label{thorem3}
The inverse of $F^U_{Y_1|Y_0}(y_1,y_0)$ given $y_0$, for any $u \in (0,1)$, is
{
\begin{align}
\label{eq34}
&F^{U-1}_{Y_1|Y_0}(u,y_0)=
F^{-1}_{Y_1}(u(F_{Y_0}(y_0)-F'_{Y_0}(y_0))+F'_{Y_0}(y_0)),
\end{align}}
and the inverse of $F^L_ {Y_1|Y_0}(y_1,y_0)$ given $y_0$ 
is
{
\begin{align}
\label{eq35}
&F^{L-1}_{Y_1|Y_0}(u,y_0)=F^{-1}_{Y_1}(1-(1-u)(F_{Y_0}(y_0)-F'_{Y_0}(y_0))-F'_{Y_0}(y_0)),
\end{align}}
where $F^{-1}_{Y_1}$ is the generalized inverse of $F_{Y_1}$.
\end{restatable}
\noindent Although $F^{U-1}_{Y_1|Y_0}(u, y_0)$ and $F^{L-1}_{Y_1|Y_0}(u, y_0)$ are not defined at points $u = 0$ and $u = 1$ in Theorem \ref{thorem3}, the uniform random variable $U \sim \text{Unif}[0,1]$ takes these values with probability zero.
Consequently, this does not pose a problem.

When $F_{Y_0}$ is continuous, i.e., $F_{Y_0}(y_0)=F'_{Y_0}(y_0)$, Eq.~\eqref{eq34} reduces to
$F^{U-1}_{Y_1|Y_0}(u,y_0)=F^{-1}_{Y_1}(F_{Y_0}(y_0))$, which is equal to the right-hand side of Eq.~\eqref{eq16}. 
In this case, our identification result under the comonotonicity assumption coincides with the identification result of \citep{Heckman1997} in Eq.~\eqref{eq-rankeq}; this aligns with Proposition~\ref{prop-equi}. 
Also, when $F_{Y_0}$ is continuous, Eq.~\eqref{eq35} reduces to 
$F^{L-1}_{Y_1|Y_0}(u,y_0)=F^{-1}_{Y_1}(1-F_{Y_0}(y_0))$, which is equal to the right-hand side of Eq.~\eqref{eq16con}. This result aligns with Theorem~\ref{theo-counter}. 



\subsection{Estimation of joint expectations over POs from finite samples}

{
We assume we are given experimental data from $\mathbb{P}(Y_x)$. 
We denote
the samples of $Y_1$ from the treatment group 
as ${\cal D}^T=\{Y^T_1,\dots,Y^T_{N^T}\}$ and the samples of $Y_0$  from the control group 
as ${\cal D}^C=\{Y^C_1,\dots,Y^C_{N^C}\}$. 
} 

{To estimate $\mathbb{E}^U[g(Y_1,Y_0)]$ and $\mathbb{E}^L[g(Y_1,Y_0)]$ from Eqs.~\eqref{eq-23} and \eqref{eq-24}, we use 
(i) i.i.d. samples from the distribution $F_{Y_0}$, which are from ${\cal D}^C$, and (ii) for each $y_0=Y^C_i$, draw samples $\tilde{Y}_i^{U,T}$ and $\tilde{Y}_i^{L,T}$ from the distributions $F^U_{Y_1|Y_0}$ and $F^L_{Y_1|Y_0}$ respectively. 
The latter samples are generated via ITS as follows: (i) Draw a sample $U_i$ from the uniform distribution $\text{Unif}[0,1]$, and (ii) Compute $\tilde{Y}_i^{U,T}$ and $\tilde{Y}_i^{L,T}$  by substituting  $u=U_i$ and $y_0=Y^C_i$ into 
 Eqs.~\eqref{eq34} and \eqref{eq35}, respectively.}
{The expressions of the empirical CDFs used in the computation are given in Appendix \ref{appC}.}
Finally, 
we use Eqs.~\eqref{eq-23} and \eqref{eq-24} to calculate the empirical expectation of $g(Y_1,Y_0)$ using generated samples $\{\tilde{Y}_i^{U,T},Y^C_i\}_{i=1}^{N^C}$ and $\{\tilde{Y}_i^{L,T},Y^C_i\}_{i=1}^{N^C}$ as follows: 
\begin{align}
\hat{\mathbb{E}}^U[g(Y_1,Y_0)]=\frac{1}{N^C}\sum_{i=1}^{N^C} g(\tilde{Y}_i^{U,T},Y^C_i), \quad \hat{\mathbb{E}}^L[g(Y_1,Y_0)]=\frac{1}{N^C}\sum_{i=1}^{N^C} g(\tilde{Y}_i^{L,T},Y^C_i).
\end{align}

These are consistent estimators as follows:
\begin{theorem}
\label{theorem4}
{Suppose that $g$ is Lipschitz continuous and that $Y_0$ and $Y_1$ have compact supports. Then $\hat{\mathbb{E}}^U[g(Y_1,Y_0)]$ and $\hat{\mathbb{E}}^L[g(Y_1,Y_0)]$ converge in probability to $\mathbb{E}^U[g(Y_1,Y_0)]$ and $\mathbb{E}^L[g(Y_1,Y_0)]$, respectively, at the rate $\mathcal{O}_p\big((N^C)^{-1/2}+(N^T)^{-1/2}\big)$.}
\end{theorem}

\subsection{Estimation from observational data \label{sec-obs}}
Our results extend immediately to settings where, instead of experimental data from $\mathbb{P}(Y_x)$, we observe only  observational data from $\mathbb{P}(X,Y)$ under the \emph{exogeneity assumption},  $Y_x\indep X$. 
Under exogeneity, we have $F_{Y_x} = F_{Y|X=x}$ and $\mathbb{P}(Y_x)=\mathbb{P}(Y|x)$. 
Consequently, the observed samples of $Y$ corresponding to $X=1$ and $X=0$  can be treated as samples from $\mathbb{P}(Y_1)$ and $\mathbb{P}(Y_0)$, respectively.

{
Because our identification and bounding results are expressed entirely in terms of $F_{Y_x}(y)$, they extend naturally to settings with observed covariates $W$ under the \emph{conditional exogeneity} assumption, $Y_x \indep X|W$. 
{We further assume positivity alongside conditional exogeneity: for each treatment level $x \in \{0,1\}$, $\mathbb{P}(X=x \mid W=w) > 0$ for almost every $w$ in the support of $W$. In finite samples, inadequate overlap may lead to unstable estimation.}
Under these assumptions, each instance of $F_{Y_x}(y)$ can be replaced by the standard covariate adjustment $\int F_{Y|X=x,W=w}(y) dF_W(w)$. Estimation of the resulting expressions can proceed via 
plug-in methods. However, developing effective estimators  when $W$ is high-dimensional remains nontrivial. }

\section{Numerical Experiments and Applications to Real-World Datasets \label{sec-real}}
The code is available at \url{https://github.com/NA679723/NeurIPS2026_Joint_Expectation.git}.

\textbf{Numerical experiments.} {We conduct numerical experiments to evaluate the finite-sample properties of our estimators. The results indicate that the estimators perform reasonably well at a sample size of  $N=100$ and are stable at $N=1000$. 
For even-order moments, our bounds are tighter than those obtained from the baseline Monte Carlo integration (MCI) of \citet{Kawakami2025_moments}.
Moreover, our estimator based on closed-form expressions is significantly more computationally efficient than the numerical optimization-based causal optimal transport (COT) estimator by \citet{Gao2025}. The details are provided in Appendix~\ref{app-exp}. }

\textbf{Applications.} {We apply our methods to two real-world datasets  ``Cholesterol Reduction'' \citep{Westfall2011} and the Lalonde dataset \citep{LaLonde1986}. We estimate the variance, skewness, and kurtosis of ICDE and ICRE. In both cases, the results reveal substantial heterogeneity in treatment effects, with high tail risk. These findings suggest caution in making treatment decisions based solely on the mean ICDE. Complete results are reported in Appendix~\ref{appE}.}



\section{Conclusion}

We study the identification and bounding 
of the joint expectation $\mathbb{E}[g(Y_1, Y_0)]$ 
over POs.  
This quantity encompasses a range of important statistical characteristics of the ICDE and the ICRE.
We establish new identification results under the comonotonicity and countercomonotonicity assumptions (Theorems~\ref{theorem2},~\ref{theo-counter}). Notably, comonotonicity  is a weaker identifying assumption  than existing assumptions in the literature (Propositions~\ref{propstr}-\ref{prop2}). 
In the absence of 
these assumptions, we present novel sharp bounds for two classes of $g$ functions (Theorem~\ref{theorem1}). These results lead to 
tighter bounds for moments of ICDE (Corollaries~\ref{cor-1}-\ref{cor-3}) and new  bounds for moments of ICRE (Corollaries~\ref{cor-4}-\ref{cor-5}).
Importantly, these results apply to any type of outcome variable $Y$—discrete, continuous, or mixed. 
Overall, our results provide new tools for investigating causal effect heterogeneity and enable analyses that were previously inaccessible.  
These results have important implications for real-world decision-making: reliance on average causal effects alone may be misleading, whereas  information on causal effect heterogeneity provides a more comprehensive and reliable basis for informed decisions by practitioners. 

{The comonotonicity and countercomonotonicity assumptions are generally not testable from observed data. While we discuss their practical plausibility (following Assumptions~\ref{comono} and~\ref{countercomono}), their credibility must ultimately be assessed based on domain knowledge and the scientific context. When these assumptions are not credible, bounding approaches offer a principled alternative. Although we obtain sharp bounds for certain moments of ICDE and ICRE, our general results are restricted to specific classes of $g$ functions. Extending sharp bounds to more general function classes is left for future work.} 

{Although our identification and bounding results extend to settings with observed covariates (Section~\ref{sec-obs}), developing effective and statistically efficient estimators in the presence of \emph{high-dimensional} covariates remains nontrivial and would require substantial additional methodological advances. We therefore leave the development of robust estimation procedures for high-dimensional covariate settings to future work. 
Another promising direction is to extend the analysis to multivariate outcomes, as considered in \citet{Torous2024}.}

\subsection*{Acknowledgments}
The authors thank the anonymous reviewers for their time and thoughtful comments. This work received no external funding. The authors declare no competing interests.

\bibliographystyle{plainnat}
\bibliography{aaai25}

@book{Pearl09,
 author = {Judea Pearl},
 title = {Causality: Models, Reasoning and Inference},
 publisher = {Cambridge University Press}, 
 year = {2009},
 edition = {2nd}
}

@article{Holland1986,
  title={Statistics and causal inference},
  author={Holland, Paul W},
  journal={Journal of the American statistical Association},
  volume={81},
  number={396},
  pages={945--960},
  year={1986},
  publisher={Taylor \& Francis}
}

@article{Kawakami2024,
      title={Probabilities of Causation for Continuous and Vector Variables}, 
  author =       {Kawakami, Yuta and Kuroki, Manabu and Tian, Jin},
        year      = {2024},
   journal = {Proceedings of the 40th Conference on Uncertainty in Artificial Intelligence (UAI-2024)},
}

@article{Tian2000,
	author = {Tian, Jin and Pearl, Judea},
	journal = {Annals of Mathematics and Artificial Intelligence},
	number = {1},
	pages = {287--313},
	title = {Probabilities of causation: Bounds and identification},
	volume = {28},
	year = {2000}}

@article{Pearl1999,
	author = {Pearl, Judea},
	journal = {Synthese},
	number = {1},
	pages = {93--149},
	title = {Probabilities Of Causation: Three Counterfactual Interpretations And Their Identification},
	volume = {121},
	year = {1999}}

@article{Gadbury2004,
	author = {Gary L. Gadbury and Hari K. Iyer and Jeffrey M. Albert},
	journal = {Journal of Statistical Planning and Inference},
	number = {2},
	pages = {163-174},
	title = {Individual treatment effects in randomized trials with binary outcomes},
	volume = {121},
	year = {2004}}

@article{Efron1979,
	author = {B. Efron},
	journal = {The Annals of Statistics},
	number = {1},
	pages = {1 -- 26},
	title = {{Bootstrap Methods: Another Look at the Jackknife}},
	volume = {7},
	year = {1979}}

@book{Hernan2024,
	author = {Hernan, M.A. and Robins, J.M.},
	publisher = {CRC Press},
	series = {Chapman \& Hall/CRC Monographs on Statistics \& Applied Probab},
	title = {Causal Inference: What If},
	year = {2024}}

@book{Westfall2011,
  title={Multiple comparisons and multiple tests using SAS},
  author={Westfall, Peter H and Tobias, Randall D and Wolfinger, Russell D},
  year={2011},
  publisher={SAS Institute}
}

@inbook{Bareinboim2022,
author = {Bareinboim, Elias and Correa, Juan D. and Ibeling, Duligur and Icard, Thomas},
title = {On Pearl’s Hierarchy and the Foundations of Causal Inference},
year = {2022},
publisher = {Association for Computing Machinery},
address = {New York, NY, USA},
edition = {1},
booktitle = {Probabilistic and Causal Inference: The Works of Judea Pearl},
pages = {507–556},
numpages = {50}
}

@article{Kendall1938,
 author = {M. G. Kendall},
 journal = {Biometrika},
 number = {1/2},
 pages = {81--93},
 publisher = {[Oxford University Press, Biometrika Trust]},
 title = {A New Measure of Rank Correlation},
 urldate = {2024-09-22},
 volume = {30},
 year = {1938}
}

@article{Greenland1999,
	author = {Sander Greenland and Judea Pearl and James M. Robins},
	journal = {Statistical Science},
	number = {1},
	pages = {29 -- 46},
	title = {{Confounding and Collapsibility in Causal Inference}},
	volume = {14},
	year = {1999}}

@article{Rosenbaum1983,
 author = {Paul R. Rosenbaum and Donald B. Rubin},
 journal = {Biometrika},
 number = {1},
 pages = {41--55},
 publisher = {[Oxford University Press, Biometrika Trust]},
 title = {The Central Role of the Propensity Score in Observational Studies for Causal Effects},
 urldate = {2024-12-31},
 volume = {70},
 year = {1983}
}

@article{Hernan2006,
	author = {Hern{\'a}n, Miguel A and Robins, James M},
	journal = {J Epidemiol Community Health},
	month = {Jul},
	number = {7},
	pages = {578--586},
	title = {Estimating causal effects from epidemiological data.},
	volume = {60},
	year = {2006}}

@article{Fay2018,
	author = {Fay, Michael P and Brittain, Erica H and Shih, Joanna H and Follmann, Dean A and Gabriel, Erin E},
	journal = {Stat Med},
	month = {Sep},
	number = {20},
	pages = {2923--2937},
	title = {Causal estimands and confidence intervals associated with Wilcoxon-Mann-Whitney tests in randomized experiments.},
	volume = {37},
	year = {2018}}

@article{Heckman1997,
	author = {Heckman, James J. and Smith, Jeffrey and Clements, Nancy},
	journal = {The Review of Economic Studies},
	month = {10},
	number = {4},
	pages = {487-535},
	title = {Making The Most Out Of Programme Evaluations and Social Experiments: Accounting For Heterogeneity in Programme Impacts},
	volume = {64},
	year = {1997}}

@inproceedings{
Wu2025,
title={Learning Counterfactual Outcomes Under Rank Preservation},
author={Peng Wu and Haoxuan Li and Chunyuan Zheng and Yan Zeng and Jiawei Chen and Yang Liu and Ruocheng Guo and Kun Zhang},
booktitle={The Thirty-ninth Annual Conference on Neural Information Processing Systems},
year={2025},
}

@article{Xie2023,
  title={Advancing Counterfactual Inference through Nonlinear Quantile Regression},
  author={Xie, Shaoan and Huang, Biwei and Gu, Bin and Liu, Tongliang and Zhang, Kun},
  journal={arXiv preprint arXiv:2306.05751},
  year={2023}
}

@article{Lu2020,
  title={Sample-efficient reinforcement learning via counterfactual-based data augmentation},
  author={Lu, Chaochao and Huang, Biwei and Wang, Ke and Hern{\'a}ndez-Lobato, Jos{\'e} Miguel and Zhang, Kun and Sch{\"o}lkopf, Bernhard},
  journal={arXiv preprint arXiv:2012.09092},
  year={2020}
}

@inproceedings{Nasr2023,
  title={Counterfactual identifiability of bijective causal models},
  author={Nasr-Esfahany, Arash and Alizadeh, Mohammad and Shah, Devavrat},
  booktitle={International Conference on Machine Learning},
  pages={25733--25754},
  year={2023},
  organization={PMLR}
}

@book{Nelsen2006,
  title={An introduction to copulas},
  author={Nelsen, Roger B},
  year={2006},
  publisher={Springer}
}

@book{Mcneil2015,
  title={Quantitative risk management: concepts, techniques and tools-revised edition},
  author={McNeil, Alexander J and Frey, R{\"u}diger and Embrechts, Paul},
  year={2015},
  publisher={Princeton university press}
}

@book{Pearl2018,
  title={The book of why: the new science of cause and effect},
  author={Pearl, Judea and Mackenzie, Dana},
  year={2018},
  publisher={Basic books}
}

@article{Kawakami2025_moments,
      title={Moments of Causal Effects}, 
   journal = {Proceedings of the 41st Conference on Uncertainty in Artificial Intelligence (UAI-2025)},
      author={Yuta Kawakami and Jin Tian},
      year={2025},
      url={https://arxiv.org/abs/2505.04971}, 
      eprint={2505.04971},
      archivePrefix={arXiv},
      primaryClass={stat.ME},
}

@article{Gabriel2024,
	author = {Gabriel, Erin E and Sachs, Michael C and Jensen, Andreas Kryger},
	journal = {Biometrika},
	month = {04},
	number = {4},
	pages = {1429-1436},
	title = {Sharp symbolic nonparametric bounds for measures of benefit in observational and imperfect randomized studies with ordinal outcomes},
	volume = {111},
	year = {2024}}

@article{Lu2019_bound,
	author = {Lu, Jiannan and Zhang, Yunshu and Ding, Peng},
	journal = {Biometrics},
	month = {11},
	number = {2},
	pages = {664-669},
	title = {Sharp Bounds on the Relative Treatment Effect for Ordinal Outcomes},
	volume = {76},
	year = {2019}}

@article{Fan2010,
 author = {Yanqin Fan and Sang Soo Park},
 journal = {Econometric Theory},
 number = {3},
 pages = {931--951},
 publisher = {Cambridge University Press},
 title = {Sharp Bounds on the Distribution of the Treatment Effect and
Their Statistical Inference},
 urldate = {2025-05-29},
 volume = {26},
 year = {2010}
}

@article{Huang2016,
	author = {Huang, Emily J. and Fang, Ethan X. and Hanley, Daniel F. and Rosenblum, Michael},
	journal = {Biostatistics},
	month = {11},
	number = {2},
	pages = {308-324},
	title = {Inequality in treatment benefits: Can we determine if a new treatment benefits the many or the few?},
	volume = {18},
	year = {2016}}

@book{Billingsley2017,
  title={Probability and measure},
  author={Billingsley, Patrick},
  year={2017},
  publisher={John Wiley \& Sons}
}

@article{Athey2006,
  title={Identification and inference in nonlinear difference-in-differences models},
  author={Athey, Susan and Imbens, Guido W},
  journal={Econometrica},
  volume={74},
  number={2},
  pages={431--497},
  year={2006},
  publisher={Wiley Online Library}
}

@article{Torous2024,
  title={An optimal transport approach to estimating causal effects via nonlinear difference-in-differences},
  author={Torous, William and Gunsilius, Florian and Rigollet, Philippe},
  journal={Journal of Causal Inference},
  volume={12},
  number={1},
  pages={20230004},
  year={2024},
  publisher={De Gruyter}
}

@InProceedings{Gao2025,
  title = 	 {Bridging Multiple Worlds: Multi-marginal Optimal Transport for Causal Partial-identification Problem},
  author =       {Gao, Zijun and Ge, Shu and Qian, Jian},
  booktitle = 	 {Proceedings of The 28th International Conference on Artificial Intelligence and Statistics},
  pages = 	 {721--729},
  year = 	 {2025},
  editor = 	 {Li, Yingzhen and Mandt, Stephan and Agrawal, Shipra and Khan, Emtiyaz},
  volume = 	 {258},
  series = 	 {Proceedings of Machine Learning Research},
  month = 	 {03--05 May},
  publisher =    {PMLR},
}

@article{Duarte2024,
  title={An automated approach to causal inference in discrete settings},
  author={Duarte, Guilherme and Finkelstein, Noam and Knox, Dean and Mummolo, Jonathan and Shpitser, Ilya},
  journal={Journal of the American Statistical Association},
  volume={119},
  number={547},
  pages={1778--1793},
  year={2024},
  publisher={Taylor \& Francis}
}

@inproceedings{Hashimoto2026,
  title={Bounds and Identification of Joint Probabilities of Potential Outcomes and Observed Variables under Monotonicity Assumptions},
  author={Hashimoto, Naoya and Kawakami, Yuta and Tian, Jin},
  booktitle={The 29th International Conference on Artificial Intelligence and Statistics (AISTATS)},
  year={2026}
}

@article{Zhang2025,
  title={Identifying and bounding the probability of necessity for causes of effects with ordinal outcomes},
  author={Zhang, Chao and Geng, Zhi and Li, Wei and Ding, Peng},
  journal={Biometrika},
  pages={asaf049},
  year={2025},
  publisher={Oxford University Press}
}

@inproceedings{Cuturi2013,
 author = {Cuturi, Marco},
 booktitle = {Advances in Neural Information Processing Systems},
 title = {Sinkhorn Distances: Lightspeed Computation of Optimal Transport},
 volume = {26},
 year = {2013}
}

@article{LaLonde1986,
  title={Evaluating the econometric evaluations of training programs with experimental data},
  author={LaLonde, Robert J},
  journal={The American economic review},
  pages={604--620},
  year={1986},
  publisher={JSTOR}
}

@article{Manski1997,
  title={Monotone treatment response},
  author={Manski, Charles},
  journal={Econometrica},
  volume={65},
  number={6},
  pages={1311--1334},
  year={1997},
  publisher={Econometric Society}
}

@inproceedings{Dawid2016,
  title={Bounding the probability of causation in mediation analysis},
  author={Dawid, A Philip and Murtas, Rossella and Musio, Monica},
  booktitle={Topics on Methodological and Applied Statistical Inference},
  pages={75--84},
  year={2016},
  organization={Springer}
}

@article{Sachs2023,
  title={A general method for deriving tight symbolic bounds on causal effects},
  author={Sachs, Michael C and Jonzon, Gustav and Sj{\"o}lander, Arvid and Gabriel, Erin E},
  journal={Journal of Computational and Graphical Statistics},
  volume={32},
  number={2},
  pages={567--576},
  year={2023},
  publisher={Taylor \& Francis}
}

@InProceedings{Zhang2022_Partial,
  title = 	 {Partial Counterfactual Identification from Observational and Experimental Data},
  author =       {Zhang, Junzhe and Tian, Jin and Bareinboim, Elias},
  booktitle = 	 {Proceedings of the 39th International Conference on Machine Learning},
  pages = 	 {26548--26558},
  year = 	 {2022},
  volume = 	 {162},
  series = 	 {Proceedings of Machine Learning Research},
  month = 	 {17--23 Jul},
  publisher =    {PMLR},
}

@article{zaffalon2024efficient,
  title={Efficient computation of counterfactual bounds},
  author={Zaffalon, Marco and Antonucci, Alessandro and Caba{\~n}as, Rafael and Huber, David and Azzimonti, Dario},
  journal={International Journal of Approximate Reasoning},
  volume={171},
  pages={109111},
  year={2024},
  publisher={Elsevier}
}

@article{Cambanis1976,
  title={Inequalities for E k (x, y) when the marginals are fixed},
  author={Cambanis, Stamatis and Simons, Gordon and Stout, William},
  journal={Zeitschrift f{\"u}r Wahrscheinlichkeitstheorie und verwandte Gebiete},
  volume={36},
  number={4},
  pages={285--294},
  year={1976},
  publisher={Springer}
}

@article{Tchen1980,
  title={Inequalities for distributions with given marginals},
  author={Tchen, Andr{\'e} H},
  journal={The Annals of Probability},
  pages={814--827},
  year={1980},
  publisher={JSTOR}
}

\newpage
\onecolumn
\appendix

\section*{\centering \LARGE Appendix to ``Identification and Bounding of Joint Expectation over Potential Outcomes"}

\section{Proof}
\label{appA}

In this appendix, we provide proofs of lemmas, theorems, and propositions in the body of the paper.

\Lemmaone*


\begin{proof}
The upper bound is attained when $(Y_1, Y_0)$ are comonotonic, while the lower bound is attained when $(Y_1, Y_0)$ are countercomonotonic \citep{Nelsen2006}. 
We have
\begin{align}
&F^L_{(Y_1,Y_0)}(y_1,y_0)=\max\{F_{Y_0}(y_0)+F_{Y_1}(y_1)-1,0\},\\
&F^U_{(Y_1,Y_0)}(y_1,y_0)=\min\{F_{Y_0}(y_0),F_{Y_1}(y_1)\}.
\end{align}
Then, Lemma \ref{lemma1} holds.
\end{proof}

\Theoremone*



\begin{proof}
Under SCM ${\cal M}$ and Assumption
\ref{comono}, we have $F_{(Y_1,Y_0)}(y_1,y_0)=F^U_{(Y_1,Y_0)}(y_1,y_0)$ since Assumption \ref{comono} implies $F_{(Y_1,Y_0)}(y_1,y_0)$ attaining the upper bound of Fr\'{e}chet–Hoeffding bounds.
Also, under SCM ${\cal M}$ and Assumption
\ref{countercomono}, we have $F_{(Y_1,Y_0)}(y_1,y_0)=F^L_{(Y_1,Y_0)}(y_1,y_0)$ since Assumption \ref{countercomono} implies $F_{(Y_1,Y_0)}(y_1,y_0)$ attaining the lower bound of Fr\'{e}chet–Hoeffding bounds.
Since $F^U_{(Y_1,Y_0)}(y_1,y_0)$ and $F^L_{(Y_1,Y_0)}(y_1,y_0)$ are identified through Lemma \ref{lemma1}, we have Theorem \ref{theorem2}.
\end{proof}

\Propositionone*


\begin{proof}
This is a corollary of Proposition \ref{propmono} since Assumption \ref{STR} implies Assumption \ref{MONO2}.
The discrete or mixed outcome is a counterexample to the converse implication.
\end{proof}

\Propositiontwo*


\begin{proof}
\noindent {\bf (Assumption \ref{comono} $\Rightarrow$ $\mathbb{P}(Y_0=1,Y_1=0)=0$ or $\mathbb{P}(Y_0=0,Y_1=1)=0$)}
{Assumption \ref{comono} implies $\mathbb{P}(Y_0=0,Y_1=1)=\max\{\mathbb{P}(Y_1=1)-\mathbb{P}(Y_0=1),0\}$ and $\mathbb{P}(Y_0=1,Y_1=0)=\max\{\mathbb{P}(Y_0=1)-\mathbb{P}(Y_1=1),0\}$.
If $\mathbb{P}(Y_1=1)\geq \mathbb{P}(Y_0=1)$, then $\mathbb{P}(Y_0=1,Y_1=0)=0$.
If $\mathbb{P}(Y_1=1)< \mathbb{P}(Y_0=1)$, then $\mathbb{P}(Y_0=0,Y_1=1)=0$.
Thus, the condition $\mathbb{P}(Y_0=1,Y_1=0)=0$ or $\mathbb{P}(Y_0=0,Y_1=1)=0$ holds.}

\noindent {\bf ($\mathbb{P}(Y_0=1,Y_1=0)=0$ or $\mathbb{P}(Y_0=0,Y_1=1)=0$ $\Rightarrow$ Assumption \ref{comono})}
When $\mathbb{P}(Y_0=1,Y_1=0)=0$,
we consider the random variable $U'$, which takes three values depending on POs as: $1=(Y_0=0,Y_1=0), 2=(Y_0=0,Y_1=1), 3=(Y_0=1,Y_1=1)$.
Then, $Y_1$ follows $\mathbb{I}(U'\geq 2)$, and $Y_0$ follows $\mathbb{I}(U'\geq 3)$ and they are monotonic increasing functions on $U'$.
Since the random variable $U'$ can be represented by a uniform distribution, then Assumption \ref{comono} holds.
{When $\mathbb{P}(Y_0=0,Y_1=1)=0$,
we consider the random variable $U'$, which takes three values depending on POs as: $1=(Y_0=0,Y_1=0), 2=(Y_0=1,Y_1=0), 3=(Y_0=1,Y_1=1)$.
Then, $Y_0$ follows $\mathbb{I}(U'\geq 2)$, and $Y_1$ follows $\mathbb{I}(U'\geq 3)$ and they are monotonic increasing functions on $U'$.
Since the random variable $U'$ can be represented by a uniform distribution, then Assumption \ref{comono} holds.}
\end{proof}

\Propositionthree*


\begin{proof}
\noindent
When $f_Y$ is monotonic increasing on $U_Y$ for all $x \in \{0,1\}$ almost surely w.r.t. $\mathbb{P}_{U_Y}$,
Assumption \ref{comono} holds since any random variable can be represented by a uniform distribution, i.e., $U_Y=F_{U_Y}^{-1}(U)$, and $F_{U_Y}^{-1}$ is monotonic increasing.
When $f_Y$ is monotonic decreasing on $U_Y$ for all $x \in \{0,1\}$ almost surely w.r.t. $\mathbb{P}_{U_Y}$, Assumption \ref{comono} also holds by considering $-U_Y$ in place of $U_Y$.

The converse, however, need not hold for a given structural representation.
For example, let $U_Y \sim \mathrm{Unif}(0,1)$ and $f_Y(x, U_Y) = |U_Y - 1/2|$ for both $x \in \{0,1\}$.
Then $Y_1 = Y_0$ almost surely, so $(Y_1, Y_0)$ is trivially comonotonic and Assumption~\ref{comono} holds.
However, $u \mapsto |u - 1/2|$ is decreasing on $(0, 1/2)$ and increasing on $(1/2, 1)$, so $f_Y(x, \cdot)$ is not monotonic on any set of full $P_{U_Y}$-measure, and Assumption~\ref{MONO2} does not hold.

Nevertheless, if $(Y_1,Y_0)$ are comonotonic, there exists $U'_Y \sim \mathrm{Unif}(0,1)$ such that $Y_x = F_{Y_x}^{-1}(U'_Y)$ for $x \in \{0,1\}$.
Setting $f'_Y(x,u) = F_{Y_x}^{-1}(u)$, which is non-decreasing in $u$, the structural function $f'_Y(X,U'_Y)$ satisfies Assumption \ref{MONO2} and $Y_x = f'_Y(x,U'_Y)$ for $x \in \{0,1\}$.
\end{proof}

\Propositionfour*

\begin{proof}
(Assumption~\ref{POD} $\Rightarrow$ Assumption~\ref{comono}) {Assumption~\ref{POD}} represents $U=F_{Y_1}(Y_1)=F_{Y_0}(Y_0)$, where $U \sim \text{Unif} [0,1]$. Since $F_{Y_1}$ and $F_{Y_0}$ are both invertible and monotonic increasing, $Y_1=F^{-1}_{Y_1}(U)$ and $Y_0=F^{-1}_{Y_0}(U)$.
Then, Assumption~\ref{comono} holds.\\
(Assumption~\ref{comono} $\Rightarrow$ Assumption~\ref{POD}) Under Assumption~\ref{comono}, we have $Y_1=F^{-1}_{Y_1}(U)$ and $Y_0=F^{-1}_{Y_0}(U)$.
Then, we have $F_{Y_1}(Y_1)=U=F_{Y_0}(Y_0)$ and Assumption~\ref{POD}.
\end{proof}

\Propositionfive*

\begin{proof}
If $\tilde\rho(Y_1,Y_0)=1$, every i.i.d. pair $(Y_1,Y_0),(Y_1',Y_0')$ is concordant and none is discordant: $(Y_1-Y_1')(Y_0-Y_0')\geq 0$ almost surely.
Since every pair of independent draws is concordant, the rank variables
$U=F_{Y_1}(Y_1)$ and $V=F_{Y_0}(Y_0)$ satisfy $U=V$ almost surely.
Then, we have $Y_0=F_{Y_0}^{-1}(U)$ and $Y_1=F_{Y_1}^{-1}(U)$, realizing both variables as non-decreasing transforms of the same latent $U$.
Thus, $(Y_1,Y_0)$ are comonotonic.

We present a counterexample showing that Assumption \ref{comono} does not imply Assumption \ref{Kendall}:
\begin{align}
\label{eq22}
\text{\bf Counterexample: }(Y_1,Y_0)=\{(1,1),(1,0),(1,0),(0,0)\}.
\end{align} 
In this case, $\tilde{\rho}_4 = 1/3$ and is not $1$.
This setting does not satisfy Assumption \ref{Kendall}.
However, our comonotonicity assumption holds.
\end{proof}

\Theoremtwo*


\begin{proof}
Under SCM ${\cal M}$ and Assumption
\ref{countercomono},
when $Y$ is continuous,
there exists a random variable $U \sim \mathrm{Unif}(0,1)$ such that:
$Y_0 = F_{Y_0}^{-1}(U)$ and $Y_1 = F_{Y_1}^{-1}(1 - U)$.
Then, we have $Y_1 = F_{Y_1}^{-1}(1 - F_{Y_0}(Y_0))$.
Then, given $Y=y_0$ and $X=0$ ($\Rightarrow Y_0=y_0$), the value of $y_1$ is identifiable by 
\begin{equation}
y_1=F^{-1}_{Y_1}(1-F_{Y_0}(y_0)).
\end{equation}
Then, the joint expectations over POs are identified by
\begin{equation}
\mathbb{E}[g(Y_1,Y_0)]=\mathbb{E}[g(F_{Y_1}^{-1}(1-F_{Y_0}(Y_0)),Y_0)].
\end{equation}
\end{proof}

\Theoremthree*





\begin{proof}
{The result is closely related to classical results on extremal couplings \citep{Cambanis1976, Tchen1980}; we include a proof for completeness.}
Let $F$ be any bivariate distribution that has the marginals, $F_{Y_1}$ and $F_{Y_0}$.
Let a function $g$ that is supermodular and bounded, and $\Omega_Y=(y^{min},y^{max})$, where $y^{min},y^{max} \in [-\infty,\infty]$.
We approximate the function $g$ as $\tilde{g}(y_1,y_0)=g(y^{min},y^{min})+\sum_{i=1}^{I}(g(a_i,y^{min})-g(a_{i-1},y^{min}))\mathbb{I}(y_1>a_i)+\sum_{j=1}^{J}(g(y^{min},b_j)-g(y^{min},b_{j-1}))\mathbb{I}(y_0>b_j)+\sum_{i=1}^{I}\sum_{j=1}^{J}(g(a_i,b_j)-g(a_{i-1},b_j)-g(a_i,b_{j-1})+g(a_{i-1},b_{j-1}))\mathbb{I}(y_1>a_i,y_0>b_j)$, which converges to $g$ as $I, J \to \infty$.
By taking expectation, $\mathbb{E}[\tilde{g}(Y_1,Y_0)]=g(y^{min},y^{min})+\sum_{i=1}^{I}(g(a_i,y^{min})-g(a_{i-1},y^{min}))\mathbb{P}(Y_1>a_i)+\sum_{j=1}^{J}(g(y^{min},b_j)-g(y^{min},b_{j-1}))\mathbb{P}(Y_0>b_j)+\sum_{i=1}^{I}\sum_{j=1}^{J}(g(a_i,b_j)-g(a_{i-1},b_j)-g(a_i,b_{j-1})+g(a_{i-1},b_{j-1}))\mathbb{P}(Y_1>a_i,Y_0>b_j)$.
Since the coefficient of $\mathbb{P}(Y_1>a_i,Y_0>b_j)$, i.e., $g(a_i,b_j)-g(a_{i-1},b_j)-g(a_i,b_{j-1})+g(a_{i-1},b_{j-1})$, is nonnegative due to the supermodularity of $g$ even for approximation, the upper bound of $\mathbb{E}[\tilde{g}(Y_1,Y_0)]$ is achieved when $Y_1$ and $Y_0$ are comonotonic, and the lower bound is achieved when $Y_1$ and $Y_0$ are countercomonotonic.
Then, taking limits $I, J \rightarrow \infty$, we have $\mathbb{E}^L[g(Y_1,Y_0)] \leq \mathbb{E}[g(Y_1,Y_0)] \leq \mathbb{E}^U[g(Y_1,Y_0)]$.
Similarly, when $g$ is submodular,
the coefficient of $\mathbb{P}(Y_1>a_i,Y_0>b_j)$, i.e., $g(a_i,b_j)-g(a_{i-1},b_j)-g(a_i,b_{j-1})+g(a_{i-1},b_{j-1})$, is nonpositive for all $i,j$.
Then, we have $\mathbb{E}^U[g(Y_1,Y_0)] \leq \mathbb{E}[g(Y_1,Y_0)] \leq \mathbb{E}^L[g(Y_1,Y_0)]$.
\end{proof}


\Lemmatwo*


\begin{proof}
We consider two random variables $(A,B)$ whose joint distribution follows {$F^U_{Y_1,Y_0}$}.
There exists a random variable $U \sim \text{Unif}(0,1)$ and non-decreasing functions $f_1$ and $f_0$ such that $A = f_1(U)$ and $B = f_0(U)$, where the marginal distributions of $A$ and $B$ are equal to the distribution of $Y$ given $X=1$ and $X=0$, respectively.
We denote $u_{y_0}=\sup\{u\in [0,1];f_0(u)\leq y_0\}$, $u_{y_1}=\sup\{u\in [0,1];f_1(u)\leq y_1\}$, and $u'_{y_0}=\sup\{u\in [0,1];f_0(u)< y_0\}$.
When $F_{Y_0}(y_0)\ne F'_{Y_0}(y_0)$, we have
\begin{align}
\mathbb{P}(A\leq y_1|B=y_0)&=\frac{\mathbb{P}(A\leq y_1,B=y_0)}{\mathbb{P}(B=y_0)}=\frac{\mathbb{P}(U\leq u_{y_1},u'_{y_0}<  U \leq u_{y_0})}{\mathbb{P}(u'_{y_0}<  U \leq u_{y_0})}\\
&=\max\left\{\frac{\min\{\mathbb{P}(U\leq u_{y_1}),\mathbb{P}(U\leq u_{y_0})\}-\mathbb{P}(U\leq  u'_{y_0})}{\mathbb{P}(U\leq u_{y_0})-\mathbb{P}(U \leq u'_{y_0})},0\right\}\\
&=\max\left\{\frac{\min\left\{F_{Y_1}(y_1),F_{Y_0}(y_0)\right\}-F'_{Y_0}(y_0)}{F_{Y_0}(y_0)-F'_{Y_0}(y_0)},0\right\}.
\end{align}
When $F_{Y_0}(y_0)= F'_{Y_0}(y_0)$, we have
\begin{align}
\mathbb{P}(A\leq y_1|B=y_0)&=\mathbb{P}(U\leq u_{y_1}|U=u_{y_0})\\
&=\mathbb{I}(u_{y_0} \leq  u_{y_1})\\
&=\mathbb{I}(F_{Y_1}(y_1)\geq F_{Y_0}(y_0))\\
&=\mathbb{I}(y_1\geq F_{Y_1}^{-1}(F_{Y_0}(y_0))).
\end{align}

We consider two random variables $(A,B)$ whose joint distribution follows $F^L_{Y_1,Y_0}$.
There exists a random variable $U \sim \text{Unif}(0,1)$ and a non-increasing function $f_0$ and a non-decreasing function $f_1$ (or a non-decreasing function $f_0$ and a non-increasing function $f_1$) such that $A=f_1(U)$ and $B=f_0(U)$, where the marginal distributions of $A$ and $B$ are equal to the distribution of $Y$ given $X=1$ and $X=0$, respectively.
We denote $u_{y_0}=\sup\{u\in [0,1];f_0(u)\leq y_0\}$, $u_{y_1}=\inf\{u\in [0,1];f_1(1-u)\leq y_1\}$, and $u'_{y_0}=\sup\{u\in [0,1];f_0(u)< y_0\}$.
When $F_{Y_0}(y_0)\ne F'_{Y_0}(y_0)$, we have
\begin{align}
\mathbb{P}(A\leq y_1|B=y_0)
&=1-\mathbb{P}(A> y_1|B=y_0)\\
&=1-\frac{\mathbb{P}(A> y_1,B=y_0)}{\mathbb{P}(B=y_0)}\\
&=1-\frac{\mathbb{P}(U\leq u_{y_1},u'_{y_0}< U \leq u_{y_0})}{\mathbb{P}(u'_{y_0}< U \leq u_{y_0})}\\
&=1-\max\left\{\frac{\min\{\mathbb{P}(U\leq u_{y_1}),\mathbb{P}(U\leq u_{y_0})\}-\mathbb{P}(U\leq  u'_{y_0})}{\mathbb{P}(U\leq u_{y_0})-\mathbb{P}(U \leq u'_{y_0})},0\right\}\\
&=1-\max\left\{\frac{\min\left\{1-F_{Y_1}(y_1),F_{Y_0}(y_0)\right\}-F'_{Y_0}(y_0)}{F_{Y_0}(y_0)-F'_{Y_0}(y_0)},0\right\}.
\end{align}
When $F_{Y_0}(y_0)= F'_{Y_0}(y_0)$, we have
\begin{align}
\mathbb{P}(A\leq y_1|B=y_0)&=\mathbb{P}(1-U\leq u_{y_1}|U=u_{y_0})\\
&=\mathbb{P}(u_{y_1} \geq 1-u_{y_0})\\
&=\mathbb{I}(F_{Y_1}(y_1)\geq 1-F_{Y_0}(y_0))\\
&=\mathbb{I}(y_1\geq F_{Y_1}^{-1}(1-F_{Y_0}(y_0))).
\end{align}
\end{proof}

\Theoremfour*


\begin{proof}
\noindent{\bf (A). When $F_{Y_0}(y_0)\ne F'_{Y_0}(y_0)$.}
If $\displaystyle\frac{\min\left\{F_{Y_1}(y_1),F_{Y_0}(y_0)\right\}-F'_{Y_0}(y_0)}{F_{Y_0}(y_0)-F'_{Y_0}(y_0)}\leq 0$, then the right-hand side of Eq.~\eqref{eq26} is equal to 0.
If $F_{Y_1}(y_1)\geq F_{Y_0}(y_0)$, then the right-hand side of Eq.~\eqref{eq26} is equal to 1.
Thus, for any $u \in(0,1)$, we have
\begin{align}
F^{U-1}_{Y_1|Y_0}(u,y_0)
&=\inf\left\{y_1;\frac{F_{Y_1}(y_1)-F'_{Y_0}(y_0)}{F_{Y_0}(y_0)-F'_{Y_0}(y_0)}\geq u\right\}\\
&=\inf\{y_1;F_{Y_1}(y_1)\geq u(F_{Y_0}(y_0)-F'_{Y_0}(y_0))+F'_{Y_0}(y_0)\}\\
&=F^{-1}_{Y_1}(u(F_{Y_0}(y_0)-F'_{Y_0}(y_0))+F'_{Y_0}(y_0)).
\end{align}
If $\displaystyle\frac{\min\left\{1-F_{Y_1}(y_1),F_{Y_0}(y_0)\right\}-F'_{Y_0}(y_0)}{F_{Y_0}(y_0)-F'_{Y_0}(y_0)}\leq 0$, then the right-hand side of Eq.~\eqref{eq27} is equal to 1.
If $1-F_{Y_1}(y_1)\geq F_{Y_0}(y_0)$, then the right-hand side of Eq.~\eqref{eq27} is equal to 0.
Thus, for any $u \in(0,1)$, we have
\begin{align}
F^{L-1}_{Y_1|Y_0}(u,y_0)&=\inf\left\{y_1;1-\frac{1-F_{Y_1}(y_1)-F'_{Y_0}(y_0)}{F_{Y_0}(y_0)-F'_{Y_0}(y_0)}\geq u\right\}\\\
&=\inf\{y_1;F_{Y_1}(y_1)\geq 1-(1-u)(F_{Y_0}(y_0)-F'_{Y_0}(y_0))-F'_{Y_0}(y_0)\}\\
&=F^{-1}_{Y_1}(1-(1-u)(F_{Y_0}(y_0)-F'_{Y_0}(y_0))-F'_{Y_0}(y_0))
\end{align}

\noindent{\bf (B). When $F_{Y_0}(y_0)= F'_{Y_0}(y_0)$.}
From Lemma \ref{lemma2},
we have
$F^{U-1}_{Y_1|Y_0}(u,y_0)=F^{-1}_{Y_1}(F_{Y_0}(y_0))$.
and
$F^{L-1}_{Y_1|Y_0}(u,y_0)=F^{-1}_{Y_1}(1-F_{Y_0}(y_0))$.
\end{proof}

\noindent{\bf Theorem \ref{theorem4}.}
{\it
{Suppose that $g$ is Lipschitz continuous and that $Y_0$ and $Y_1$ have compact supports. Then $\hat{\mathbb{E}}^U[g(Y_1,Y_0)]$ and $\hat{\mathbb{E}}^L[g(Y_1,Y_0)]$ converge in probability to $\mathbb{E}^U[g(Y_1,Y_0)]$ and $\mathbb{E}^L[g(Y_1,Y_0)]$, respectively, at the rate $\mathcal{O}_p({N^C}^{-1/2}+{N^T}^{-1/2})$.}
}
\begin{proof}
We prove the claim for $\hat{\mathbb{E}}^U$; the proof for $\hat{\mathbb{E}}^L$ is identical.
Under comonotonicity, $\mathbb{E}^U[g(Y_1,Y_0)] = \int_0^1 g(F_{Y_1}^{-1}(u),F_{Y_0}^{-1}(u))\,du$.
The error of $\hat{\mathbb{E}}^U$ splits into two parts.

\textit{(i) Sampling error from $U_i$.}
Conditional on the data, each generated pair $(\tilde Y_i^{U,T}, Y_i^C)$ equals $(\hat F_{Y_1}^{-1}(V_i), \hat F_{Y_0}^{-1}(V_i))$ for a uniform draw $V_i$ within the rank interval of $Y_i^C$.
Hence the conditional mean of $\hat{\mathbb{E}}^U$ is $\int_0^1 g(\hat F_{Y_1}^{-1}(u),\hat F_{Y_0}^{-1}(u))\,du$.
Since $g$ is bounded, its conditional variance is at most $\|g\|_\infty^2/N^C$, so this error is $\mathcal{O}_p({N^C}^{-1/2})$.

\textit{(ii) Error from estimating the marginals.}
Since $g$ is Lipschitz, the remaining error is bounded by a constant times $\sum_{x\in\{0,1\}}\int_0^1|\hat F_{Y_x}^{-1}(u)-F_{Y_x}^{-1}(u)|\,du$.
Because $Y_x$ has compact support $[a_x,b_x]$, each integral is at most $(b_x-a_x)\sup_y|\hat F_{Y_x}(y)-F_{Y_x}(y)|$.
This sup-norm error is $\mathcal{O}_p({N^T}^{-1/2})$ for $x=1$ and $\mathcal{O}_p({N^C}^{-1/2})$ for $x=0$ by the Dvoretzky--Kiefer--Wolfowitz inequality.

Combining (i) and (ii) gives the rate $\mathcal{O}_p({N^C}^{-1/2}+{N^T}^{-1/2})$.
\end{proof}

\section{Additional Bounding Corollaries}
\label{appB}

The $m$-th central moments of ICDE and ICRE are given as follows:
\begin{align}
&\mathbb{E}[\{(Y_1-Y_0)-(\mathbb{E}[Y_1]-\mathbb{E}[Y_0])\}^m]\\
&=\sum_{l=0}^m \binom{m}{l} (-1)^l \mathbb{E}[(Y_1-Y_0)^{m-l}] \left( \mathbb{E}[Y_1-Y_0] \right)^l\\
&=\sum_{l=0}^m \sum_{k=0}^{m-l} \binom{m}{l}  \binom{m-l}{k}(-1)^{l+k}\mathbb{E}[Y_1^{m-l-k}Y_0^{k}]\left( \mathbb{E}[Y_1-Y_0] \right)^l,
\end{align}
\begin{align}
&\mathbb{E}[\{(Y_1/Y_0)-(\mathbb{E}[Y_1/Y_0])\}^m]=\sum_{l=0}^m \binom{m}{l} (-1)^l \mathbb{E}[(Y_1/Y_0)^{m-l}] \left( \mathbb{E}[Y_1/Y_0] \right)^l.
\end{align}
Then, we have the following corollaries:
\begin{corollary}[The odd-order central moments of ICDE]
Under SCM ${\cal M}$,
if $Y_1$ and $Y_0$ take non-negative values,
the odd-order central moments of the ICDE are bounded by
\begin{equation}
\begin{aligned}
&\sum_{l=0}^m \sum_{k=0}^{m-l} \binom{m}{l} \binom{m-l}{k}\Big(\mathbb{I}((-1)^{l+k}\left( \mathbb{E}[Y_1-Y_0] \right)^l\leq 0)(-1)^{l+k}\left( \mathbb{E}[Y_1-Y_0] \right)^l\mathbb{E}^U[Y_1^{m-l-k}Y_0^{k}]\\
&+\mathbb{I}((-1)^{l+k}\left( \mathbb{E}[Y_1-Y_0] \right)^l> 0)(-1)^{l+k}\left( \mathbb{E}[Y_1-Y_0] \right)^l\mathbb{E}^L[Y_1^{m-l-k}Y_0^{k}]\Big)\\
&\leq \mathbb{E}[\{(Y_1-Y_0)-(\mathbb{E}[Y_1]-\mathbb{E}[Y_0])\}^m]\\
&\leq \sum_{l=0}^m \sum_{k=0}^{m-l} \binom{m}{l} \binom{m-l}{k}\Big(\mathbb{I}((-1)^{l+k}\left( \mathbb{E}[Y_1-Y_0] \right)^l\leq 0)(-1)^{l+k}\left( \mathbb{E}[Y_1-Y_0] \right)^l\mathbb{E}^L[Y_1^{m-l-k}Y_0^{k}]\\
&+\mathbb{I}((-1)^{l+k}\left( \mathbb{E}[Y_1-Y_0] \right)^l> 0)(-1)^{l+k}\left( \mathbb{E}[Y_1-Y_0] \right)^l\mathbb{E}^U[Y_1^{m-l-k}Y_0^{k}]\Big).
\end{aligned}
\end{equation}
\end{corollary}

Similarly, the skewness and kurtosis of ICDE and ICRE are also bounded.
We denote the lower and upper bounds of variance, the central third and fourth moments of ICDE (or ICRE) as $\overline{\mu}_L^{(2)}$, $\overline{\mu}_U^{(2)}$, $\overline{\mu}_L^{(3)}$, $\overline{\mu}_U^{(3)}$, $\overline{\mu}_L^{(4)}$, and $\overline{\mu}_U^{(4)}$.
We can calculate the bounds of the skewness of ICDE (or ICRE) as follows:
\begin{align}
\frac{\overline{\mu}_L^{(3)}}{\big(\overline{\mu}_U^{(2)}\big)^{3/2}}\mathbb{I}\big(\overline{\mu}_L^{(3)}\geq 0\big)+\frac{\overline{\mu}_L^{(3)}}{\big(\overline{\mu}_L^{(2)}\big)^{3/2}}\mathbb{I}\big(\overline{\mu}_L^{(3)}< 0\big)
&\leq \frac{\overline{\mu}^{(3)}}{\big(\overline{\mu}^{(2)}\big)^{3/2}} \nonumber\\
&\leq \frac{\overline{\mu}_U^{(3)}}{\big(\overline{\mu}_L^{(2)}\big)^{3/2}}\mathbb{I}\big(\overline{\mu}_U^{(3)}\geq 0\big)+\frac{\overline{\mu}_U^{(3)}}{\big(\overline{\mu}_U^{(2)}\big)^{3/2}}\mathbb{I}\big(\overline{\mu}_U^{(3)}< 0\big),
\end{align}
and the bounds of the kurtosis of ICDE (or ICRE) are calculated as follows:
\begin{align}
\frac{\overline{\mu}_L^{(4)}}{\big(\overline{\mu}_U^{(2)}\big)^{2}}\leq \frac{\overline{\mu}^{(4)}}{\big(\overline{\mu}^{(2)}\big)^{2}} \leq\frac{\overline{\mu}_U^{(4)}}{\big(\overline{\mu}_L^{(2)}\big)^{2}}.
\end{align}
These bounds are not sharp.

\section{Details of Estimation of Joint Expectation over POs}
\label{appC}
 The following empirical CDFs will be used in the computation:
\begin{align}
\hat{F}_{Y_0}(y_0)&=\frac{1}{N^C}\sum\nolimits_{j=1}^{N^C}\mathbb{I}(Y_j^C\leq y_0),\\
\hat{F}'_{Y_0}(y_0)&=\frac{1}{N^C}\sum\nolimits_{j=1}^{N^C}\mathbb{I}(Y_j^C< y_0), \\
\hat{F}_{Y_1}(y_1)&=\frac{1}{N^T}\sum\nolimits_{i=1}^{N^T}\mathbb{I}(Y_i^T\leq y_1),\\
\hat{F}^{-1}_{Y_1}(p) &= \min_{i=1,\dots,N^T}\{Y^T_i;\hat{F}_{Y_1}(Y^T_i)\geq p\}.
\end{align}
\noindent 
We have 
\begin{align}
&\tilde{Y}_i^{U,T}=\hat{F}^{-1}_{Y_1}(U_i(\hat{F}_{Y_0}(Y^C_i)-\hat{F}'_{Y_0}(Y^C_i))+\hat{F}'_{Y_0}(Y^C_i)),\\
&\tilde{Y}_i^{L,T}=\hat{F}^{-1}_{Y_1}(1-(1-U_i)(\hat{F}_{Y_0}(Y^C_i)-\hat{F}'_{Y_0}(Y^C_i))-\hat{F}'_{Y_0}(Y^C_i)).
\end{align}

\section{Numerical Experiments \label{app-exp}}

We illustrate the finite-sample properties of our estimators.
We pick up the moments of ICDE and ICRE as examples. 

\subsection{Experiments for Expectations over ICDE}

\noindent {\bf Baselines.}
We compare our estimators with the following baselines.\\
(i) The estimator based on the rank preservation assumptions (RP) (Assumptions \ref{POD} and \ref{Kendall}) \citep{Heckman1997,Wu2025}. 
RP corresponds to setting $u=0$ in our estimator for $\mathbb{E}^U[g(Y_1,Y_0)]$ instead of drawing $u$ from $\text{Unif}(0,1)$; at $u=0$, the right-hand side of Eq.~\eqref{eq34} becomes $F^{-1}_{Y_1}(F'_{Y_0}(y_0))$, which coincides with that of Eq.~\eqref{eq16} when $F_{Y_0}$ is continuous (i.e., $F_{Y_0}=F'_{Y_0}$).\\
(ii) The Monte Carlo integration-based estimator (MCI) in \citep{Kawakami2025_moments}.
The MCI estimator requires Assumption \ref{MONO2} {for identification and can also compute bounds.} It requires specifying  the outcome domain range, $\Omega_Y=[a,b]$, which we used the minimum and maximum values in the dataset.\\ 
(iii) The causal-optimal-transportation-based estimator (COT) for causal bounding in \citep{Gao2025}. COT relies on linear programming.\\
(iv) COT with Sinkhorn regularization. Sinkhorn regularization is an entropic regularization technique for optimal transport that enables efficient computation.


\noindent {\bf Settings.}
We assume the following SCMs (A) and (B):
\begin{align}
&\text{(A): } Y:=-2U_Y\mathbb{I}(X=1)-U_Y\mathbb{I}(U_Y<0,X=0),\\
&\text{(B): } Y:=4(U_Y)^2\mathbb{I}(X=1)-(U_Y)^2\mathbb{I}(U_Y>0,X=0){+1},
\end{align}
where $U_Y\sim \text{Unif}(-1,1)$ and $X \sim \text{Bern}({0.5})$.
$\text{Bern}(p)$ represents a Bernoulli distribution with probability $p$.
$Y_1$ is a continuous variable, whereas $Y_0$ is a mixed variable under both SCM (A) and (B).
$Y$ takes nonnegative values under SCM (B).
Both SCMs satisfy the exogeneity assumption,
and violate  the strictly monotonicity (Assumption \ref{STR}), the rank preservation  (Assumption \ref{POD}), and the rank preservation  on Kendall's $\tilde{\rho}$ (Assumption \ref{Kendall}) assumptions.
SCM (A) satisfies the comonotonicity (Assumption \ref{comono}) and  the monotonicity over $f_Y$ (Assumption \ref{MONO2}) whereas SCM (B) does not. 
In setting (A), we focus on point identification. In setting (B), we compute bounds. 
We simulate 100 times with the total sample size $N^C+N^T=20,100,1000$, respectively,   
{and report the
mean along with the 95\% confidence intervals (CI) for each estimate.}
The experiments are performed using an Apple M4 (24GB).

\begin{table}[!tb]
\centering
\caption{{Mean estimates with 95\% CI of moments (M.) of ICDE under SCM (A).} 
{The estimates reported as "Ours" are obtained under the comonotonicity assumption, which the SCM satisfies.}
}
\label{tab:ne1}
\scalebox{0.8}{
\begin{tabular}{c|cccc}
\hline
Statistics & $N=20$ & $N=100$ & $N=1000$ &  Ground Truth \\
\hline
\hline
2nd M. (\textbf{Ours})  & $1.033 [0.134,2.609]$ &  $0.869 [0.501,1.189]$ & $0.836[0.698,0.984]$ &$0.833$ \\
2nd M. (RP)  & $2.081 [0.867,4.144]$ &  $2.036 [1.271,3.040]$ & $2.137[1.843,2.418]$ &$0.833$ \\
2nd M. (MCI)  &$0.890[0.333,1.542]$ &  $0.882 [0.577,1.192]$ & $0.843[0.716,0.980]$ &$0.833$  \\
\hline
3rd M. (\textbf{Ours})   &  $-0.924 [-4.586,0.873]$ & $-0.817 [-1.817,-0.058]$ & $-0.903 [-1.162,-0.657]$ &$-0.875$ \\
3rd M. (RP) & $-3.896 [-10.148,-0.862]$ &  $-3.648 [-5.662,-2.321]$ & $-3.820[-4.355,-3.428]$ &$-0.875$ \\
3rd M. (MCI)  &$-0.665 [-2.337,0.569]$ & $-0.806 [-1.361,-0.237]$ & $-0.888 [-1.114,-0.637]$ &$-0.875$ \\
\hline
4th M. (\textbf{Ours})   &  $2.125 [0.026,5.906]$ & $1.813 [0.427,4.022]$& $1.734 [1.303,2.263]$ & $1.700$  \\
4th M. (RP)  & $8.436 [1.166,32.412]$ &  $7.451 [4.375,10.298]$ & $8.026[6.905,8.973]$ &$1.700$  \\
4th M. (MCI)  & $1.824 [0.213,3.884]$ & $1.687 [0.952,2.883]$& $1.685 [1.218,2.447]$ &  $1.700$  \\
\hline
\end{tabular}
}
\end{table}

\begin{table}[!tb]
\centering
\caption{Mean estimates with 95\% CIs for the upper (UB) and lower (LB) bounds of the moments (M.) of ICDE under SCM (B).
A checkmark ($\checkmark$) indicates that the estimator is based on sharp bounds.
COT is a numerical approach and does not use closed-form expressions for the bounds.
The Sinkhorn regularization parameter is 100.
COT results for $N=1000$ are omitted because of computational resource constraints.
}
\label{tab:ne2}
\scalebox{0.75}{
\begin{tabular}{c|rrrr}
\hline
Statistics & $N=20$ & $N=100$ & $N=1000$ & True M. value \\
\hline
\hline
UB of 2nd M. (\textbf{Ours}) $\checkmark$ & $4.393 [1.602, 7.648]$ & $4.342 [3.180, 5.742]$ & $4.347 [3.958, 4.783]$ & $4.100$ \\
LB of 2nd M. (\textbf{Ours}) $\checkmark$ & $3.656 [0.912, 6.574]$ & $3.382 [2.079, 4.807]$ & $3.359 [2.899, 3.809]$ & $4.100$ \\
UB of 2nd M. (MCI) $\checkmark$ & $4.359 [1.585, 7.557]$ & $4.374 [3.154, 5.870]$ & $4.348 [3.831, 4.822]$ & $4.100$ \\
LB of 2nd M. (MCI) & $0.984 [0.053, 4.734]$ & $0.642 [0.232, 1.342]$ & $0.572 [0.394, 0.772]$ & $4.100$ \\
UB of 2nd M. (COT) & $4.420 [2.460, 7.815]$ & $4.260 [3.073, 5.519]$ & -- & $4.100$ \\
LB of 2nd M. (COT) & $3.634 [1.813, 6.705]$ & $3.319 [2.253, 4.475]$ & -- & $4.100$ \\
UB of 2nd M. (COT; Sinkhorn) & $3.986 [2.078, 7.202]$ & $3.714 [2.578, 4.948]$ & $3.748 [3.342, 4.164]$ & $4.100$ \\
LB of 2nd M. (COT; Sinkhorn) & $3.980 [2.071, 7.189]$ & $3.706 [2.572, 4.938]$ & $3.740 [3.335, 4.155]$ & $4.100$ \\
\hline
UB of 3rd M. (\textbf{Ours}) & $16.261 [4.434, 32.235]$ & $16.501 [11.386, 22.740]$ & $17.358 [15.527, 19.211]$ & $13.500$ \\
LB of 3rd M. (\textbf{Ours}) & $8.466 [-0.247, 22.496]$ & $6.951 [3.084, 11.547]$ & $7.349 [6.007, 8.695]$ & $13.500$ \\
UB of 3rd M. (MCI) & $14.467 [3.791, 29.534]$ & $14.895 [9.526, 20.092]$ & $15.309 [13.789, 17.401]$ & $13.500$ \\
LB of 3rd M. (MCI) & $0.933 [0.000, 7.114]$ & $0.163 [0.000, 0.637]$ & $0.106 [0.009, 0.259]$ & $13.500$ \\
UB of 3rd M. (COT) & $14.244 [3.196, 26.965]$ & $15.296 [9.470, 23.007]$ & -- & $13.500$ \\
LB of 3rd M. (COT) & $9.363 [1.254, 17.763]$ & $9.310 [5.436, 13.841]$ & -- & $13.500$ \\
UB of 3rd M. (COT; Sinkhorn) & $11.474 [3.244, 23.400]$ & $11.592 [6.942, 18.224]$ & $11.450 [9.996, 12.907]$ & $13.500$ \\
LB of 3rd M. (COT; Sinkhorn) & $11.222 [3.165, 22.961]$ & $11.249 [6.681, 17.709]$ & $11.108 [9.715, 12.544]$ & $13.500$ \\
\hline
UB of 4th M. (\textbf{Ours}) $\checkmark$ & $49.656 [4.782, 121.923]$ & $57.574 [30.873, 84.743]$ & $58.409 [50.830, 66.497]$ & $48.944$ \\
LB of 4th M. (\textbf{Ours}) $\checkmark$ & $27.307 [1.071, 61.272]$ & $28.798 [11.022, 49.474]$ & $27.877 [21.527, 34.861]$ & $48.944$ \\
UB of 4th M. (MCI) $\checkmark$ & $50.697 [6.849, 106.510]$ & $59.138 [29.202, 89.327]$ & $59.425 [51.387, 67.160]$ & $48.944$ \\
LB of 4th M. (MCI) & $0.329 [0.000, 2.484]$ & $0.026 [0.000, 0.278]$ & $0.010 [0.000, 0.137]$ & $48.944$ \\
UB of 4th M. (COT) & $52.338 [11.757, 111.960]$ & $57.791 [30.035, 95.728]$ & -- & $48.944$ \\
LB of 4th M. (COT) & $29.518 [3.053, 68.991]$ & $29.173 [15.233, 53.132]$ & -- & $48.944$ \\
UB of 4th M. (COT; Sinkhorn) & $37.118 [3.098, 99.440]$ & $42.780 [25.074, 71.245]$ & $43.257 [36.782, 50.880]$ & $48.944$ \\
LB of 4th M. (COT; Sinkhorn) & $30.258 [2.921, 81.995]$ & $33.744 [19.564, 54.594]$ & $33.707 [28.965, 39.999]$ & $48.944$ \\
\hline
\end{tabular}
}
\end{table}

\noindent {\bf Results.}
Table \ref{tab:ne1} reports {the  mean estimates with 95\% CI} of the second, third, and fourth moments (2nd M., 3rd M., and 4th M.) of the ICDE for each estimator under SCM (A), together with the corresponding ground truths.
{Both our estimator and MCI perform well (their identification assumptions are satisfied in SCM (A)), while RP exhibits severe bias (its identification assumption is violated).}

Table \ref{tab:ne2} reports the estimates of the upper (UB) and lower (LB) bounds  of the second, third, and fourth moments of the ICDE for each estimator under SCM (B), together with the corresponding 
{true moment values}.
The true values fall within the estimated bounds. 
For the 2nd and 4th moments, our bounds are tighter than the MCI bounds, particularly for the lower bounds.
This is consistent with our theoretical results that for the even-order moments of ICDE, our bounds are sharp, while MCI's upper bounds are sharp but lower bounds are not.
For the 3rd moment, neither method's bounds are sharp, and our upper bound can be wider than that of MCI.


\begin{table}[!tb]
\centering
\caption{Runtime Comparison for a Single Simulation.
COT for $N^C+N^T = 2000$ and $5000$ exceeds the available memory. The regularization parameter of Sinkhorn is set to 100.}
\label{tab:time1}
\scalebox{1}{
\begin{tabular}{c|cccccc}
\hline
Time & $N^C+N^T=20$ & $100$ & $1000$ & $2000$ & $5000$  \\
\hline
\textbf{Ours} & 0.051 sec. &  0.055 sec. & 0.134 sec.& 0.344 sec.& 1.110 sec.  \\
COT & 0.001 sec.&  0.243 sec. & 203.476 sec. & - & - \\
COT (Sinkhorn) & 0.001 sec.&  0.003 sec. & 0.155 sec. & 0.782 sec. & 8.230 sec. \\
\hline
\end{tabular}
}
\end{table}

Table \ref{tab:time1} reports the runtime for a single simulation of our estimator and the COT estimator.
COT relies on linear programming, which can be computationally demanding, $\mathcal{O}((N^T+N^C)^3)$ time and $\mathcal{O}((N^T+N^C)^2)$ memory.
In contrast, our estimator requires 
$\mathcal{O}((N^T+N^C) \log (N^T+N^C))$ time
and $\mathcal{O}(N^T+N^C)$ memory.
COT estimators are very similar to ours for $N = 20$ and $N = 100$. However, results for $N = 1000$ could not be obtained; nonetheless, our estimator significantly reduces computational cost while achieving similar confidence intervals compared to COT.
This reduction in computational resources arises from our closed-form expressions for the bounds.
Regularization methods for COT, such as Sinkhorn \citep{Cuturi2013}, greatly reduce computational cost; however, they can introduce substantial bias in the estimates and may fail to cover the ground truth, as shown in Table \ref{tab:ne2}.


\subsection{Experiments for Expectations over ICRE}


\noindent {\bf Baseline.}
We compare our estimators with (i) the RP estimator and (ii) the causal-optimal-transportation (COT) estimator. 
MCI does not work for ICRE. 

\noindent {\bf Settings.}
We assume the following SCM (C) and (D):
\begin{align}
&\text{(C): } Y:=6U_Y\mathbb{I}(X=1)+U_Y\mathbb{I}(U_Y>0.75,X=0)+2, U_Y\sim \text{Unif}(0,1),\\
&\text{(D): } Y:=6(U_Y)^2\mathbb{I}(X=1)+(U_Y)^2
\mathbb{I}(U_Y>0.75,X=0)+2, U_Y\sim \text{Unif}(-1,1),
\end{align}
where $X \sim \text{Bern}({0.5})$. $Y_1$ is a continuous variable, whereas $Y_0$ is a mixed variable under both SCM (C) and (D).
$Y$ takes positive values under both (C) and (D). 
Both SCMs satisfy the exogeneity,
and violate the rank preservation (Assumption \ref{POD}) and the rank preservation  on Kendall's $\tilde{\rho}$ (Assumption \ref{Kendall}).
SCM (C) satisfies the comonotonicity (Assumption \ref{comono}), whereas SCM (D) does not.
In setting (C), we focus on point identification. In setting (D), we compute bounds.
We simulate 100 times with the sample size $N^C+N^T=20,100,1000$, respectively.

\noindent {\bf Results.}
Table \ref{tab:ne3} reports {the  mean estimates with 95\% CI} of the first, second, third, and fourth moments (1st M., 2nd M., 3rd M., and 4th M.) of the ICRE for each estimator under SCM (C), together with the corresponding ground truths.
{Our estimator performs well, while RP exhibits severe bias (its identification assumption is violated).}

Table \ref{tab:ne4} reports the estimates of the bounds of the first, second, third, and fourth moments of the ICRE  under SCM (D), together with the corresponding 
{true moment values}. 
These bounds are theoretically sharp. 

    
Similarly to ICDE, COT estimators are very similar to ours for $N = 20$ and $N = 100$. However, results for $N = 1000$ could not be obtained due to computational constraints. Regularization methods for COT, such as Sinkhorn \citep{Cuturi2013}, can greatly reduce computational cost; however, they may introduce substantial bias and fail to cover the ground truth, as shown in Table \ref{tab:ne4}.

\begin{table}[!tb]
\centering
\caption{{Mean estimates with 95\% CI of moments (M.) of} ICRE under SCM (C).
{The estimates reported as "Ours" are obtained under the comonotonicity assumption, which the SCM satisfies.}}
\label{tab:ne3}
\scalebox{0.85}{
\begin{tabular}{c|cccc}
\hline
Statistics & $N=20$ & $N=100$ & $N=1000$ &  Ground Truth \\
\hline
1st M. (\textbf{Ours}) & $2.173[1.591,2.794]$ &  $2.212 [1.942,2.527]$ & $2.231[2.144,2.342]$ &$2.223$ \\
1st M. (RP) & $1.523 [1.154,1.999]$ &  $1.417 [1.293,1.584]$ & $1.378[1.337,1.429]$ &$2.223$ \\
\hline
2nd M. (\textbf{Ours}) & $5.615 [2.891,9.210]$ &  $5.425 [4.171,6.554]$ & $5.318[4.825,5.896]$ &$5.292$ \\
2nd M. (RP) & $2.578[1.418,4.289]$ &  $2.341 [1.842,2.849]$ & $2.335[2.197,2.507]$ &$5.292$ \\
\hline
3rd M. (\textbf{Ours})& $14.252 [6.384,26.138]$ &$13.392 [8.298,18.450]$ &  $13.233 [11.594,15.262]$ &$13.227$ \\
3rd M. (RP)& $4.808 [1.763,8.546]$ &$4.675 [3.507,5.853]$ &  $4.759 [4.267,5.228]$ &$13.227$ \\
\hline
4th M. (\textbf{Ours})& $45.743[9.299,106.961]$ &$34.699 [20.084,54.742]$ &  $34.826 [29.713,42.228]$ &$34.519$\\
4th M. (RP)& $9.529 [1.264,21.594]$ &$10.152 [6.984,13.769]$ &  $10.922 [9.998,11.896]$ &$34.519$\\
\hline
\end{tabular}
}
\end{table}

\begin{table}[!tb]
\centering
\caption{{Mean estimates with 95\% CI of upper (UB) and lower (LB) bounds of moments (M.) of} ICRE under SCM (D). A checkmark ($\checkmark$) indicates whether the estimator is based on sharp bounds (for our method).\\
We do not report COT results for $N = 1000$ due to computational resource constraints.
The regularization parameter of Sinkhorn is set to 100.}
\label{tab:ne4}
\scalebox{0.75}{
\begin{tabular}{c|cccc}
\hline
Statistics & $N=20$ & $N=100$ & $N=1000$ &  {True M. value} \\
\hline
\hline
UB of 1st M. (\textbf{Ours}) ($\checkmark$)  & $1.985 [1.417,2.839]$ &$1.985 [1.727,2.275]$ &  $1.959 [1.867,1.965]$ & $1.884$ \\
LB of 1st M. (\textbf{Ours}) ($\checkmark$)  & $1.875 [1.386,2.631]$ &$1.874 [1.644,2.193]$ &  $1.873 [1.789,2.196]$ &$1.884$ \\

UB of 1st M. (COT) &$1.993 [1.454, 2.528]$    &    $1.965 [1.709, 2.247]$ & - &$1.884$  \\
LB of 1st M. (COT)  & $1.912 [1.355, 2.451]$   &     $1.875 [1.598, 2.166]$ & -&$1.884$ \\

UB of 1st M. (COT; Sinkhorn)  &$1.913 [1.449, 2.444]$    &    $1.926 [1.700, 2.175]$   &     $1.937 [1.870, 1.999]$&$1.884$  \\
LB of 1st M. (COT; Sinkhorn)  & $1.913 [1.449, 2.444]$   &     $1.926 [1.700, 2.175]$   &     $1.937 [1.870, 1.999]$&$1.884$ \\
\hline

UB of 2nd M. (\textbf{Ours}) ($\checkmark$)  & $4.572 [1.824,8.662]$ &$4.736 [3.382,6.263]$ &  $4.770 [4.322,5.256]$ & $4.124$ \\
LB of 2nd M. (\textbf{Ours}) ($\checkmark$)  & $4.077 [1.667,8.637]$ &$4.002 [2.980,5.415]$ &  $4.021 [3.600,4.445]$ &$4.124$ \\

UB of 2nd M. (COT)  &$4.519 [2.425, 7.374]$    &    $4.746 [3.557, 6.085]$  & - &$4.124$  \\
LB of 2nd M. (COT)  & $4.017 [2.065, 6.696]$    &    $3.986 [2.995, 5.401]$ &- &$4.124$ \\

UB of 2nd M. (COT; Sinkhorn)   &$4.404 [2.210, 6.406]$ &       $4.553 [3.666, 5.742]$   &     $4.548 [4.184, 4.914]$ &$4.124$  \\
LB of 2nd M. (COT; Sinkhorn)   & $4.397 [2.208, 6.391]$  &      $4.544 [3.657, 5.728]$  &      $4.539 [4.176, 4.906]$ &$4.124$ \\

\hline
UB of 3rd M. (\textbf{Ours}) ($\checkmark$)  & $13.111 [3.629,22.645]$ &$13.447 [7.029,20.343]$ &  $12.915 [10.515,14.949]$ &$10.251$ \\
LB of 3rd M. (\textbf{Ours}) ($\checkmark$) & $9.578[2.926,25.494]$ &$9.900 [5.703,15.309]$ &  $9.338 [7.772,10.831]$ &$10.251$ \\

UB of  3rd  M. (COT)  &$12.530 [3.934, 22.876]$  &    $13.375 [9.299, 17.724]$  & - &$10.251$  \\
LB of  3rd  M. (COT)   &$9.918 [3.263, 20.067]$   &    $9.752 [6.556, 13.875]$ &  -&$10.251$ \\

UB of  3rd  M. (COT; Sinkhorn)  &$12.282 [4.443, 20.953]$ &     $12.183 [8.529, 16.304]$ &     $12.362 [11.011, 13.879]$  &$10.251$  \\
LB of  3rd  M. (COT; Sinkhorn)   &$12.110 [4.414, 20.933]$ &     $11.977 [8.397, 16.067]$ &     $12.144 [10.807, 13.670]$&$10.251$ \\
\hline
UB of 4th M. (\textbf{Ours})($\checkmark$)  & $39.128 [5.569,100.039]$ &$40.786 [22.117,65.312]$ &  $39.770 [33.264,48.310]$ &$28.132$ \\
LB of 4th M. (\textbf{Ours}) ($\checkmark$)  & $27.363 [5.082,80.050]$ &$25.602 [12.818,45.459]$ &  $23.712 [19.036,29.0653]$ &$28.132$ \\

UB of  4th  M. (COT)  &$37.648 [11.288, 67.875]$   &  $38.829 [24.517, 55.395]$ & -&$28.132$  \\
LB of  4th  M. (COT)  & $27.073 [7.380, 64.114]$   &   $24.258 [13.207, 38.913]$ & - &$28.132$\\

UB of  4th  M. (COT; Sinkhorn)  &$37.120 [12.387, 72.289]$ &    $37.963 [22.942, 52.821]$  &   $38.267 [32.890, 43.705]$ &$28.132$  \\
LB of  4th  M. (COT; Sinkhorn)  & $33.442 [11.887, 68.291]$  &   $33.860 [20.124, 48.042]$ &    $34.026 [29.083, 38.746] $&$28.132$\\
\hline
\end{tabular}
}
\end{table}

\subsection{Experiments Under the Rank Preservation Assumption}

We provide additional numerical experiments.

\noindent {\bf Settings.}
We assume the following SCMs (E) and (F):
\begin{align}
&\text{(E): } Y:=-(X+1)U_Y,
U_Y\sim \text{Unif}(-1,1),\\
&\text{(F): } Y:=X+1+U_Y,
U_Y\sim \text{Unif}(0,1).
\end{align}
These SCMs satisfy the rank preservation assumption (Assumption \ref{POD}).
In settings (E) and (F), we compare our estimator with RP and MCI under the assumption that all estimators are unbiased.

\noindent{\bf Results.}
Table \ref{tab:appne1} shows the results of the numerical experiments for ICDE under SCM (E) and Table \ref{tab:appne2} shows the results of the numerical experiments for ICRE under SCM (F).
All mean estimates are close to the ground truth, particularly when the sample size is large.
Our estimator, RP, and MCI have almost the same 95\% CIs.

\begin{table}[tb]
\centering
\caption{Results of numerical experiments for ICDE under SCM (E).
{The estimates reported as "Ours" are obtained under the comonotonicity assumption, which the SCM satisfies.}}
\label{tab:appne1}
\scalebox{0.85}{
\begin{tabular}{c|cccc}
\hline
Statistics & $N=20$ & $N=100$ & $N=1000$ &  Ground Truth \\
\hline
\hline
2nd M. (\textbf{Ours})  & $0.627 [0.177,1.631]$ &  $0.380 [0.203,0.632]$ & $0.338[0.274,0.397]$ &$0.333$ \\
2nd M. (RP)  & $0.664 [0.176,1.536]$ &  $0.407 [0.208,0.651]$ & $0.343[0.286,0.415]$ &$0.333$ \\
2nd M. (MCI)  &$0.616[0.139,1.205]$ &  $0.382 [0.198,0.610]$ & $0.338[0.265,0.407]$ &$0.333$  \\
\hline
3rd M. (\textbf{Ours})   &  $-0.038 [-1.116,1.013]$ & $-0.000 [-0.289,0.359]$ & $-0.000 [-0.106,0.075]$ &$0.000$ \\
3rd M. (RP) & $-0.403 [-1.866,0.641]$ &  $-0.040 [-0.280,0.255]$ & $-0.008[-0.093,0.066]$ &$0.000$ \\
3rd M. (MCI)  &$0.002 [-1.070,1.176]$ & $0.023 [-0.345,0.295]$ & $-0.001 [-0.112,0.099]$ &$0.000$ \\
\hline
4th M. (\textbf{Ours})   &  $0.789 [0.047,2.714]$ & $0.263 [0.106,0.496]$& $0.205 [0.153,0.272]$ & $0.200$  \\
4th M. (RP)  & $0.774 [0.035,2.275]$ &  $0.272 [0.105,0.601]$ & $0.206[0.151,0.262]$ &$0.200$  \\
4th M. (MCI)  & $0.870 [0.054,4.082]$ & $0.269 [0.086,0.556]$& $0.205 [0.080,0.336]$ &  $0.200$  \\
\hline
\end{tabular}
}
\end{table}

\begin{table}[tb]
\centering
\caption{Results of numerical experiments for ICRE under SCM (F).
{The estimates reported as "Ours" are obtained under the comonotonicity assumption, which the SCM satisfies.}}
\label{tab:appne2}
\scalebox{0.85}{
\begin{tabular}{c|cccc}
\hline
Statistics & $N=20$ & $N=100$ & $N=1000$ &  Ground Truth \\
\hline
1st M. (\textbf{Ours}) & $1.703[1.473,1.926]$ &  $1.693 [1.586,1.797]$ & $1.696[1.668,1.728]$ &$1.693$ \\
1st M. (RP) & $1.685 [1.446,1.927]$ &  $1.695 [1.607,1.798]$ & $1.695[1.664,1.723]$ &$1.693$ \\
\hline
2nd M. (\textbf{Ours}) & $2.977 [2.114,3.910]$ &  $2.867 [2.508,3.259]$ & $2.883[2.773,2.981]$ &$2.886$ \\
2nd M. (RP) & $2.821 [2.139,3.790]$ &  $2.883 [2.526,3.283]$ & $2.884[2.783,2.998]$ &$2.886$ \\
\hline
3rd M. (\textbf{Ours})& $5.233 [3.395,7.428]$ &$5.020 [4.227,5.981]$ &  $4.967 [4.692,5.315]$ &$4.961$ \\
3rd M. (RP)& $4.904 [2.953,7.017]$ &$4.983 [4.121,5.962]$ &  $4.978 [4.632,5.252]$ &$4.961$ \\
\hline
4th M. (\textbf{Ours})& $8.726[4.956,13.613]$ &$8.806 [7.016,10.951]$ &  $8.595 [8.010,9.305]$ &$8.575$\\
4th M. (RP)& $8.312 [5.237,12.498]$ &$8.592 [6.513,10.973]$ &  $8.607 [8.004,9.308]$ &$8.575$\\
\hline
\end{tabular}
}
\end{table}

\section{Applications to Real-World Datasets}
\label{appE}

\subsection{Cholesterol Reduction Dataset}

\noindent{\bf Dataset.}
We analyze a medical dataset ``Cholesterol Reduction'' \citep{Westfall2011} (\texttt{https://search.r-project.org/CRAN/refmans/multcomp/html/cholestero\\l.html}). 
This study investigates the effect of three formulations of the same drug ($X$) on cholesterol reduction ($Y$).
The three formulations consist of 20 mg taken once daily (“1 time”, $X=1$), 10 mg taken twice daily (“2 times”, $X=2$), and 5 mg taken four times daily (“4 times”, $X=4$).
The dataset has a sample size of 10 for each treatment group.
The domain of $Y$ is bounded within $[5,19]$, and all values are strictly positive.
The variance, skewness, and kurtosis of ICDE for this dataset were analyzed by \citet{Kawakami2025_moments}.

We assume the exogeneity condition holds, as \citet{Westfall2011} reported no 
potential confounding variables.
We evaluate the variance, skewness, and kurtosis of the ICDE for the effects from $X = 1$ to $X = 2$, from $X = 2$ to $X = 4$, and from $X = 1$ to $X = 4$, respectively, and compare the results with those reported in \citep{Kawakami2025_moments}. 
In addition, we assess the mean, variance, skewness, and kurtosis of the ICRE for the same pairs of treatment levels.
{Comonotonicity is plausible when, at the same total exposure, splitting the dose does not materially alter efficacy profiles, so that patients who respond well under one dosing schedule also respond well under the other. For example, comonotonicity of $(Y_1, Y_2)$ means that patients who would experience a larger cholesterol reduction on 20 mg once daily would also experience a larger reduction on 10 mg twice daily.}
We conduct the bootstrapping \citep{Efron1979} to provide the means and 95\% confidence intervals (CIs) for each estimator.

\noindent{\bf Results.}
The complete results are reported in Table \ref{tab:app1}.
For ICDE,  our method and MCI agree on the estimated variance, but differ substantially on the estimated  skewness and kurtosis. Their skewness and kurtosis estimates are likely unreliable given their wide CIs. In general, our estimates have significantly narrower  CIs than MCI. Still, these CIs are relatively wide due to the small number of samples. 
If these mean estimates are to be trusted, 
{under the comonotonicity assumption,} the estimated skewness indicates that the distribution of the ICDE is approximately symmetric, meaning that the number of individuals with effects greater than the average is roughly equal to the number of individuals with effects less than the average, in contrast to the large positive skewness reported by \citet{Kawakami2025_moments}. The estimated kurtosis suggests heavier tails than the Gaussian distribution (kurtosis = 3), albeit to a lesser extent than that reported by \citet{Kawakami2025_moments}, which indicates a high number of outliers with causal
effects that differ significantly from the average. 
{Given these results {under the comonotonicity assumption}, practitioners may hesitate to implement the treatment  based solely on the mean ICDE, as doing so could result in substantially positive or negative impacts that deviate from the mean for a meaningful minority of individuals.}

The  bound for the variance (and skewness and kurtosis) of ICDE is quite wide, indicating that the results obtained under  comonotonicity  could be misleading if, in fact, comonotonicity does not hold.


{Under the comonotonicity assumption,}
the estimated mean of ICRE indicates a positive effect (greater than 1), {which is consistent with the positive mean of ICDE. The estimated bound for the mean of ICRE is  relatively narrow, and confirms the positive effect without assuming comonotonicity.} 
The estimated variance suggests that the heterogeneity of the ICRE is relatively limited.
{The estimated skewness values indicate that the ICRE distributions are approximately symmetric. The kurtosis estimates suggest lighter tails than those of a Gaussian distribution for $Y_2/Y_1$ and $Y_4/Y_1$, but slightly heavier tails for $Y_4/Y_2$.}




We provide further information about the application to the cholesterol reduction data.
We evaluate the variance, skewness, and kurtosis of the ICDE for the effects from $X = 1$ to $X = 2$, from $X = 2$ to $X = 4$, and from $X = 1$ to $X = 4$, respectively, and compare the results with those reported in \citet{Kawakami2025_moments}. 
In addition, we assess the mean, variance, skewness, and kurtosis of the ICRE for the same pairs of treatment levels.

Table \ref{tab:app1} shows all the results for ICDE $Y_2-Y_1$ and ICRE $Y_2/Y_1$,
Table \ref{tab:app2} shows the results for ICDE $Y_4-Y_2$ and ICRE $Y_4/Y_2$,
and Table \ref{tab:app3} shows the results for ICDE $Y_4-Y_1$ and ICRE $Y_4/Y_1$.
The results of MCI were reported by \citep{Kawakami2025_moments}.
{Under the comonotonicity assumption, the estimates suggest the following.
The ICDE $Y_4 - Y_2$ exhibits a similar mean and variance to $Y_2 - Y_1$, but is slightly negatively skewed and displays heavier tails.
The ICDE $Y_4 - Y_1$ exhibits a larger mean and smaller variance compared to $Y_2 - Y_1$, with an approximately symmetric distribution and relatively lighter tails.
The ICRE $Y_4/Y_2$ exhibits a smaller mean and variance compared to $Y_2/Y_1$, but shows greater skewness and heavier tails.
The ICRE $Y_4/Y_1$ exhibits a greater mean and variance compared to $Y_2/Y_1$, with slightly lower skewness and slightly lighter tails.
All bounds for skewness and kurtosis are considerably wide.}

{These conclusions, which rely on the comonotonicity assumption, may change if the potential outcomes have a different dependence structure.}

\begin{table}[tb]
    \centering
    \caption{Estimates of the statistics for $Y_2-Y_1$ and $Y_2/Y_1$.
   We report the means of the estimates along with their 95\% CIs.
   The results of MCI were reported by \citep{Kawakami2025_moments}.
   MCI is not applicable to ICRE.
   A checkmark ($\checkmark$) indicates whether the estimator is based on sharp bounds (for our method and MCI).}
    \label{tab:app1}
\scalebox{1}{
    \begin{tabular}{l|c}
    \hline
    Statistics for $Y_2-Y_1$ and $Y_2/Y_1$  &  Estimates\\
        \hline\hline
        Mean of ICDE & 3.432 [0.914, 6.104] \\
        \hline
    Variance of ICDE {under comonotonicity} (Ours) & 2.920 [0.551,6.978] \\
   Variance of ICDE {under monotonicity over $f_Y$} (MCI) & 3.072 [0.297,8.610] \\
   \hline
  Skewness of ICDE {under comonotonicity}  (Ours) & -0.184 [-2.612,2.713] \\
   Skewness of ICDE  {under monotonicity over $f_Y$} (MCI) & 21.027 [-6.747,34.504]\\
   \hline
   Kurtosis of ICDE {under comonotonicity}  (Ours) & 6.332 [1.264,15.078] \\
   Kurtosis of ICDE  {under monotonicity over $f_Y$} (MCI) & 21.312 [0.000,203.885] \\
            \hline\hline
UB of Variance of ICDE (Ours) ($\checkmark$) & 30.508 [11.422,54.639] \\
   LB of Variance of ICDE (Ours) ($\checkmark$) & 2.920 [0.551,6.978] \\
   UB of Variance of ICDE (MCI) ($\checkmark$)  & 29.851 [12.173, 53.740] \\
   LB of Variance of ICDE (MCI) & 0.000 [0.000,0.000] \\
      UB of Variance of ICDE (COT)  & 29.732 [13.317, 47.979] \\
   LB of Variance of ICDE (COT)& 2.858 [0.873, 6.150] \\
   \hline
   UB of Skewness of ICDE (Ours) & 5.580 [3.020,11.677] \\
   LB of Skewness of ICDE (Ours) & -1.204 [-6.722,15.312] \\
   UB of Skewness of ICDE (MCI) & 71296.725 [0.000,258849.124] \\
   LB of Skewness of ICDE (MCI) & -13168.940 [-64712.250,0.000]\\
   
   UB of Skewness of ICDE (COT) & 29.057 [1.762, 90.612] \\
   LB of Skewness of ICDE (COT) & -27.294 [-98.295, -2.509]\\
   \hline
   UB of Kurtosis of ICDE (Ours) & 155.876 [73.433,298.713] \\
   LB of Kurtosis of ICDE (Ours) & 0.000 [0.000,0.000] \\
   UB of Kurtosis of ICDE (MCI) &6.638$\times10^{6}$ [0.000, 3.526$\times10^{7}$] \\
   LB of Kurtosis of ICDE (MCI) &  0.000 [0.000,0.000] \\
   UB of Kurtosis of ICDE (COT) &473.925 [31.378, 1817.645] \\
   LB of Kurtosis of ICDE (COT) &  0.031 [0.002, 0.141] \\
      \hline\hline
   Mean of ICRE {under comonotonicity}  & 1.765 [1.210,2.442] \\
   Variance of ICRE {under comonotonicity}  & 0.267 [0.041,0.577] \\
   Skewness of ICRE {under comonotonicity}  & 
   {0.378[-0.916,1.778]}
   \\
   Kurtosis of ICRE {under comonotonicity}  & 
   {2.328[1.282,4.785]}\\
      \hline \hline
   UB of Mean of ICRE (Ours) ($\checkmark$) & 2.314 [1.559,3.215] \\
   LB of Mean of ICRE (Ours)  ($\checkmark$) & 1.765 [1.210,2.442] \\
   UB of Mean of ICRE (COT) & 2.271 [1.479, 3.071] \\
   LB of Mean of ICRE (COT) & 1.741 [1.174, 2.508] \\
    \hline
   UB of Variance of ICRE (Ours)  & 5.540 [2.059,9.854] \\
   LB of Variance of ICRE (Ours)  & 0.000 [0.000,0.000] \\
   UB of Variance of ICRE (COT)  & 5.258 [1.541, 9.517]\\
   LB of Variance of ICRE (COT)  & 0.000 [0.000,0.000] \\
    \hline
   UB of Skewness of ICRE (Ours)  & 56382.41 [6343.13,111289.8] \\
   LB of Skewness of ICRE (Ours)  & -50319.46 [-102465.4,-5773.89] \\
   UB of Skewness of ICRE (COT)  & 1.101$\times10^{13}$ [9.568$\times10^{11}$ , 2.567$\times10^{13}$] \\
   LB of Skewness of ICRE (COT)  & -7.236$\times10^{10}$ [-4.137$\times10^{11}$, 0.048]\\
    \hline
   UB of Kurtosis of ICRE (Ours)  & 488100.6 [39205.61,1430178] \\
   LB of Kurtosis of ICRE (Ours)  & 0.000 [0.000,0.000] \\
   UB of Kurtosis of ICRE (COT)  & 4.829$\times10^{13}$ [2.407$\times10^{12}$, 1.214$\times10^{14}$] \\
   LB of Kurtosis of ICRE (COT)  &  0.013 [0.000, 0.074] \\
      \hline
    \end{tabular}
    }
\end{table}

\begin{table}[tb]
    \centering    \caption{Estimates of the statistics for $Y_4-Y_2$ and $Y_4/Y_2$.
   We report the means of the estimates along with their 95\% CIs.
   The results of MCI were reported by \citep{Kawakami2025_moments}.
   MCI is not applicable to ICRE.
   A checkmark ($\checkmark$) indicates whether the estimator is based on sharp bounds (for our method and MCI).}
    \label{tab:app2}
\scalebox{1}{
    \begin{tabular}{l|c}
    \hline
    Statistics for $Y_4-Y_2$ and $Y_4/Y_2$ &  Estimates\\
        \hline\hline
        Mean of ICDE & 3.124 [0.364, 5.820] \\
        \hline
    Variance of ICDE {under comonotonicity}  (Ours) & 2.891 [0.613,7.143] \\
   Variance of ICDE  {under monotonicity over $f_Y$} (MCI) & 3.104 [0.297, 8.610] \\
   \hline
  Skewness of ICDE {under comonotonicity}  (Ours) & -0.762 [-6.468,2.089] \\
   Skewness of ICDE  {under monotonicity over $f_Y$} (MCI) & -3.373 [-28.287, 8.821]\\
   \hline
   Kurtosis of ICDE {under comonotonicity}  (Ours) &8.737 [1.661,23.059] \\
   Kurtosis of ICDE  {under monotonicity over $f_Y$} (MCI) &  18.765 [0.000, 203.983] \\
                \hline\hline
UB of Variance of ICDE (Ours) ($\checkmark$) & 30.508 [9.285,51.361] \\
   LB of Variance of ICDE (Ours) ($\checkmark$) & 2.891 [0.613,7.143] \\
   UB of Variance of ICDE (MCI) ($\checkmark$)  & 29.503 [10.532, 61.318] \\
   LB of Variance of ICDE (MCI) & 0.000 [0.000,0.000] \\
   UB of Variance of ICDE (COT)   & 30.965 [13.196, 53.903] \\
   LB of Variance of ICDE (COT) & 3.373 [0.824, 9.508] \\
   \hline
   UB of Skewness of ICDE (Ours) & 7.487 [3.984,14.582] \\
   LB of Skewness of ICDE (Ours) & -4.051 [-10.147,-0.0718] \\
   UB of Skewness of ICDE (MCI) & 34944.621 [0.000, 129424.512]\\
   LB of Skewness of ICDE (MCI) & -79272.510 [-268151.400, 0.000] \\
   UB of Skewness of ICDE (COT) & 32.322 [1.246, 131.101]\\
   LB of Skewness of ICDE (COT) & -33.339 [-143.721, -1.297] \\
   \hline
   UB of Kurtosis of ICDE (Ours) & 259.648 [166.984,430.6644] \\
   LB of Kurtosis of ICDE (Ours) & 0.000 [0.000,0.000] \\
   UB of Kurtosis of ICDE (MCI) & 3.262$\times 10^{6}$ [0.000, 2.226$\times 10^{7}$]  \\
   LB of Kurtosis of ICDE (MCI) & 0.000 [0.000,0.000]\\
   UB of Kurtosis of ICDE (COT) & 665.578 [28.824, 3107.350] \\
   LB of Kurtosis of ICDE (COT) & 0.000 [0.000,0.000]\\
      \hline \hline
   Mean of ICRE {under comonotonicity}  & 1.446 [1.166,1.879] \\
   Variance of ICRE {under comonotonicity}  & 0.151 [0.010,0.441] \\
   Skewness of ICRE {under comonotonicity}  & 
   {0.776[-1.083,2.089]}\\
   Kurtosis of ICRE {under comonotonicity}  & 
   {3.600[1.502,7.317]}\\
      \hline \hline
   UB of Mean of ICRE (Ours) ($\checkmark$) & 1.655 [1.244,2.263] \\
   LB of Mean of ICRE (Ours) ($\checkmark$) & 1.446 [1.166,1.879] \\
      UB of Mean of ICRE (COT) & 1.676 [1.227, 2.249] \\
   LB of Mean of ICRE (COT)  & 1.461 [1.127, 1.913] \\
    \hline
   UB of Variance of ICRE (Ours) & 1.735 [0.244,4.392] \\
   LB of Variance of ICRE (Ours) & 0.000 [0.000,0.000] \\
   UB of Variance of ICRE (COT) & 1.883 [0.369, 4.472] \\
   LB of Variance of ICRE (COT) & 0.000 [0.000,0.000] \\
    \hline
   UB of Skewness of ICRE (Ours) & 15346.36 [1207.602,38685.48] \\
   LB of Skewness of ICRE (Ours) & -13089.57 [-33060.27,-1196.08] \\
   UB of Skewness of ICRE (COT) & 3.019$\times 10^{12}$ [4.039$\times 10^{10}$, 9.351$\times 10^{12}$] \\
   LB of Skewness of ICRE (COT) & -1.781$\times 10^{10}$ [-8.175$\times 10^{10}$, 0.148] \\
    \hline
   UB of Kurtosis of ICRE (Ours) & 83329.95 [1994.425,239344.1] \\
   LB of Kurtosis of ICRE (Ours) & 0.000 [0.000,0.000] \\
   UB of Kurtosis of ICRE (COT) & 1.026$\times 10^{13}$ [4.790$\times 10^{10}$, 3.433$\times 10^{13}$] \\
   LB of Kurtosis of ICRE (COT) & 0.041 [0.001, 0.254] \\
      \hline
    \end{tabular}
    }
\end{table}

\begin{table}[tb]
    \centering
    \caption{Estimates of the statistics for $Y_4-Y_1$ and $Y_4/Y_1$.
   We report the means of the estimates along with their 95\% CIs.
    The results of MCI were reported by \citep{Kawakami2025_moments}.
    MCI is not applicable to ICRE.
    A checkmark ($\checkmark$) indicates whether the estimator is based on sharp bounds (for our method and MCI).}
    \label{tab:app3}
    \scalebox{1}{
    \begin{tabular}{l|c}
    \hline
    Statistics for $Y_4-Y_1$ and $Y_4/Y_1$ &  Estimates\\
        \hline\hline
        Mean of ICDE & 6.551 [4.549, 9.034] \\
        \hline
    Variance of ICDE {under comonotonicity}  (Ours) & 1.700 [0.286,3.287] \\
   Variance of ICDE  {under monotonicity over $f_Y$} (MCI) & 2.078 [0.000, 5.953] \\
   \hline
  Skewness of ICDE {under comonotonicity}  (Ours) & 0.004 [-2.323,1.860] \\
   Skewness of ICDE  {under monotonicity over $f_Y$} (MCI) & 5.797 [-24.688, 65.521] \\
   \hline
   Kurtosis of ICDE {under comonotonicity}  (Ours) & 4.338 [1.155,11.044] \\
   Kurtosis of ICDE  {under monotonicity over $f_Y$} (MCI) & 14.338 [0.000, 242.5153] \\
          \hline\hline
UB of Variance of ICDE (Ours) ($\checkmark$) & 24.353 [12.546,42.149] \\
   LB of Variance of ICDE (Ours) ($\checkmark$) & 1.700 [0.286,3.287] \\
   UB of Variance of ICDE (MCI) ($\checkmark$)  & 25.160 [9.034, 42.910] \\
   LB of Variance of ICDE (MCI) & 0.000 [0.000,0.000] \\
   
   UB of Variance of ICDE (COT)  & 25.949 [14.145, 43.069]\\
   LB of Variance of ICDE (COT) & 2.152 [0.486, 5.057]\\
   \hline
   UB of Skewness of ICDE (Ours) & 58.819 [-3.814,375.223] \\
   LB of Skewness of ICDE (Ours) & 10.499 [6.123,17.745] \\
   UB of Skewness of ICDE (MCI) & 34135.710 [0.000, 113246.400] \\
   LB of Skewness of ICDE (MCI) & -19251.891 [-97068.380, 0.000] \\
   
   UB of Skewness of ICDE (COT) & 36.701 [1.288, 140.127] \\
   LB of Skewness of ICDE (COT) & -36.297 [-136.666, -3.202]\\
   \hline
   UB of Kurtosis of ICDE (Ours) & 265.288 [164.641,436.681] \\
   LB of Kurtosis of ICDE (Ours) & 0.000 [0.000,0.000] \\
   UB of Kurtosis of ICDE (MCI) &  2.292$\times10^{6}$  [0.000, 1.763$\times10^{7}$] \\
   LB of Kurtosis of ICDE (MCI) & 0.000 [0.000,0.000] \\
   UB of Kurtosis of ICDE (COT) &  664.012 [43.686, 2992.521] \\
   LB of Kurtosis of ICDE (COT) & 0.024 [0.001, 0.121]\\
      \hline\hline
   Mean of ICRE {under comonotonicity}  (Ours) & 2.554 [1.777,3.681] \\
   Variance of ICRE {under comonotonicity}  (Ours) & 0.640 [0.099,1.902] \\
   Skewness of ICRE {under comonotonicity}  (Ours) & 
   {0.294[-0.777,1.377]}\\
   Kurtosis of ICRE {under comonotonicity}  (Ours) &
   {2.209[1.229,4.653]}\\
      \hline\hline
   UB of Mean of ICRE (Ours) ($\checkmark$) & 3.028 [2.087,4.164] \\
   LB of Mean of ICRE (Ours) ($\checkmark$) & 2.554 [1.777,3.681] \\
      UB of Mean of ICRE (COT) & 2.974 [2.039, 3.994] \\
   LB of Mean of ICRE (COT) & 2.502 [1.772, 3.462] \\
   \hline
   UB of Variance of ICRE (Ours) & 6.540 [2.534,12.705] \\
   LB of Variance of ICRE (Ours) & 0.019 [0.000,0.157] \\
   
   UB of Variance of ICRE (COT) & 6.454 [2.351, 11.704]\\
   LB of Variance of ICRE (COT) & 0.000 [0.000, 0.000] \\
   \hline
   UB of Skewness of ICRE (Ours) & 94626 [17344.85,199385.5] \\
   LB of Skewness of ICRE (Ours) & -87341.18 [-191424.8,-16479.25] \\
   
   UB of Skewness of ICRE (COT) & 1.201$\times10^{13}$ [1.645$\times10^{12}$, 2.910$\times10^{13}$] \\
   LB of Skewness of ICRE (COT) &  -2.376$\times10^{11}$ [-1.557$\times10^{12}$, 0.051]\\
   \hline
   UB of Kurtosis of ICRE (Ours) & 882022.3 [218388.7,2450092] \\
   LB of Kurtosis of ICRE (Ours) & 0.000 [0.000,0.000] \\
   UB of Kurtosis of ICRE (COT) &  5.752$\times10^{13}$ [4.749$\times10^{12}$, 1.541$\times10^{14}$]\\
   LB of Kurtosis of ICRE (COT) & 0.026 [0.001, 0.110] \\
      \hline
    \end{tabular}
    }
\end{table}

\subsection{Lalonde Dataset}

\noindent{\bf Dataset.}
We analyze the Lalonde dataset \citep{LaLonde1986} (\texttt{\url{https://users.nber.org/~rdehejia/nswdata2.html}}).
This dataset contains both experimental and observational data and is commonly used as a benchmark to assess whether observational inference methods can recover results consistent with experimental findings.
In contrast to this line of research, in this paper, we evaluate the shape of the distribution of causal effects, which cannot be directly computed even from experimental data.
The sample size of this dataset is 722.
The treatment variable is a binary indicator (1 = treated, 0 = control). 
We use real earnings in 1978 as the outcome variable, which ranges from 0 to 6.031$\times 10^{4}$.
Most individuals have positive, continuous income, while 196 subjects report zero income; thus, the outcome is mixed. 
For the calculation of the ICRE statistic, we add 1 to the outcome to avoid division by zero.
We select the regularization parameter by increasing it as $1, 10, 100, \ldots$ , and stop when the computation finishes within one hour.
We do not report COT results due to computational constraints.
We conduct the bootstrapping \citep{Efron1979} to provide the means and 95\% CIs for each estimator.

We assume the exogeneity condition holds, as we use experimental data  from the National Supported Work (NSW) demonstration \citep{LaLonde1986}.
We evaluate the variance, skewness, and kurtosis of the ICDE, and the mean, variance, skewness, and kurtosis of the ICRE for $Y_1 - Y_0$ and $Y_1/Y_0$, respectively.
Comonotonicity of $(Y_0, Y_1)$ means that individuals who would earn more under the control condition would also earn more under the treatment condition. 
This is plausible if the job training program benefits participants in proportion to their pre-existing earning potential, so that individuals who would perform better without the program also perform better with it. 
Comonotonicity would be less plausible if the program were designed as a remedial intervention that disproportionately benefits the lowest earners, since in that case individuals who rank low under control might rank higher under treatment.


\noindent{\bf Results.}
The complete results are also reported in Table \ref{tab:app1_lalonde}.
The mean of ICDE is positive, indicating that the treatment has a positive effect on average.
Researchers may conclude that the treatment should be implemented.
However, statistics based on higher moments of ICDE reveal risks associated with the treatment.
The variance of ICDE is non-negligible relative to the scale of the outcome $[0, 6.031\times 10^{4}]$.
This implies substantial heterogeneity in the causal effect of the treatment on earnings.
Furthermore, the distribution is positively skewed and highly kurtotic, which together imply the presence of substantial tail risk, including both large upside gains and potential downside losses for some individuals.

The mean of ICRE is large, suggesting that the treatment increases the outcome by a factor of about 50 on average.
However, similarly to ICDE, ICRE exhibits substantial heterogeneity in causal effects.
ICRE is positively skewed and highly kurtotic.
ICRE is also highly kurtotic.
They suggest the presence of tail risks in ICRE.

We also report results under the rank preservation assumption. 
These results differ from our estimates, particularly for the ICRE statistics, suggesting that imposing rank preservation in the presence of mixed variables can introduce bias.

Similarly, COT with Sinkhorn regularization may exhibit substantial regularization bias.
Such biases can lead to severely misleading scientific conclusions.


\begin{table}[tb]
    \centering    \caption{Estimates of the statistics for $Y_1-Y_0$ and $Y_1/Y_0$.
   We report the means of the estimates along with their 95\% CIs.
   MCI is not applicable to ICRE.
   A checkmark ($\checkmark$) indicates whether the estimator is based on sharp bounds (for our method and MCI).
The regularization parameter of Sinkhorn is set to $10^9$.
   }
    \label{tab:app1_lalonde}
\scalebox{1}{
    \begin{tabular}{l|c}
    \hline
    Statistics for $Y_1-Y_0$ and $Y_1/Y_0$  &  Estimates\\
        \hline\hline
        Mean of ICDE  &  840.603 [-261.218,1782.534] \\
    Variance of ICDE {under comonotonicity} (Ours)  & 4.299$\times10^{6}$ [4.688$\times10^{5}$ ,1.236$\times10^{7}$] \\
        Variance of ICDE (RP)  & 2.929$\times10^{6}$ [2.993$\times10^{5}$ ,1.049$\times10^{7}$] \\
  Skewness of ICDE {under comonotonicity} (Ours) & 5.903 [-2.422 ,11.735] \\
    Skewness of ICDE (RP) & 4.983 [-7.903,13.555]\\
   Kurtosis of ICDE {under comonotonicity} (Ours)  &  76.837[5.483,204.682] \\
      Kurtosis of ICDE (RP)  &  63.362 [4.905,227.056]\\
            \hline\hline
UB of Variance of ICDE (Ours)($\checkmark$) & 1.288$\times10^{8}$ [1.007$\times10^{8}$,1.680$\times10^{8}$]  \\
   LB of Variance of ICDE (Ours) ($\checkmark$) & 4.299$\times10^{6}$ [4.688$\times10^{5}$ ,1.236$\times10^{7}$]\\

   UB of Variance of ICDE (MCI)($\checkmark$) & 1.851$\times10^{8}$ [1.446$\times10^{8}$, 2.383$\times10^{8}$]  \\
   LB of Variance of ICDE (MCI) & 0.000 [0.000 ,0.000]\\

   UB of Variance of ICDE (COT; Sinkhorn) & 1.282$\times10^{8}$  [9.665$\times10^{7}$ , 1.760$\times10^{8}$ ]  \\
   LB of Variance of ICDE (COT; Sinkhorn) & 1.018$\times10^{8}$  [8.113$\times10^{7}$, 1.331$\times10^{8}$]\\
   UB of Skewness of ICDE (Ours) &  1.923$\times 10^{3}$ [96.144,1.015$\times 10^{4}$] \\
   LB of Skewness of ICDE (Ours)  &  -1.734$\times 10^{3}$ [-1.001$\times 10^{4}$,-43.391]\\

   UB of Skewness of ICDE (MCI) & 7.864$\times 10^{16}$ [3.100$\times 10^{16}$, 1.676$\times 10^{17}$]  \\
   LB of Skewness of ICDE (MCI)  &  -5.458$\times 10^{16}$[-7.178$\times 10^{16}$, -4.009$\times 10^{16}$] \\

    UB of Skewness of ICDE (COT; Sinkhorn) &  -0.841 [-1.488, -0.144] \\
   LB of Skewness of ICDE (COT; Sinkhorn)  &  -3.756 [-6.349, -0.913]\\
   UB of Kurtosis of ICDE (Ours)  &  1.056$\times 10^{8}$ [6.861$\times 10^{6}$,4.086$\times 10^{8}$] \\
   LB of Kurtosis of ICDE (Ours)  & 0.000 [0.000,0.000] \\

   UB of Kurtosis of ICDE (MCI)  &  1.318$\times 10^{23}$ [5.100$\times 10^{22}$, 2.844$\times 10^{23}$]  \\
   LB of Kurtosis of ICDE (MCI)  & 0.000 [0.000,0.000] \\

      UB of Kurtosis of ICDE (COT; Sinkhorn)  &  20.641 [6.582, 30.991] \\
   LB of Kurtosis of ICDE (COT; Sinkhorn)  & 0.473 [1.295$\times 10^{-4}$, 1.040] \\
      \hline\hline
   Mean of ICRE {under comonotonicity} (Ours) &  50.782 [5.957,139.223] \\
      Mean of ICRE (RP) &  1.353[1.054,1.680]\\
   Variance of ICRE {under comonotonicity} (Ours) &  4.903$\times 10^{3}$ [2.898$\times 10^{3}$ ,2.221$\times 10^{5}$ ] \\
      Variance of ICRE (RP) &  2.420 [0.041,9.0247] \\
   Skewness of ICRE {under comonotonicity} (Ours) &  4.497 [2.741,8.634] \\
      Skewness of ICRE (RP) &  9.415 [4.685,14.411] \\
   Kurtosis of ICRE {under comonotonicity} (Ours) &  29.678 [9.915,89.070]\\
   Kurtosis of ICRE (RP) & 100.260 [25.990,202.785]\\
      \hline\hline
   UB of Mean of ICRE (Ours) ($\checkmark$) &  4.271$\times 10^{3}$ [3.532$\times 10^{3}$,5.044$\times 10^{3}$] \\
   LB of Mean of ICRE (Ours) ($\checkmark$) & 50.782 [5.957,139.223] \\

   UB of Mean of ICRE (COT; Sinkhorn) &  1.496$\times 10^{3}$ [1.194$\times 10^{3}$, 1.863$\times 10^{3}$] \\
   LB of Mean of ICRE (COT; Sinkhorn) & 1.496$\times 10^{3}$ [1.194$\times 10^{3}$, 1.863$\times 10^{3}$] \\

   UB of Variance of ICRE (Ours) &  5.523$\times 10^{7}$ [3.488$\times 10^{7}$ ,8.883$\times 10^{7}$ ]\\
   LB of Variance of ICRE (Ours) &  4.903$\times 10^{3}$ [2.898$\times 10^{3}$ ,2.221$\times 10^{5}$ ] \\

      UB of Variance of ICRE (COT; Sinkhorn) &  4.338$\times 10^{7}$ [1.844$\times 10^{7}$, 8.878$\times 10^{7}$]\\
   LB of Variance of ICRE (COT; Sinkhorn) &  1.302$\times 10^{7}$ [1.007$\times 10^{7}$, 1.622$\times 10^{7}$] \\

   UB of Skewness of ICRE (Ours) &  2.180$\times 10^{15}$ [9.447$\times 10^{14}$,5.022$\times 10^{15}$] \\
   LB of Skewness of ICRE (Ours) &  -1.058$\times 10^{15}$ [-1.986$\times 10^{15}$,-5.929$\times 10^{14}$] \\

      UB of Skewness of ICRE (COT; Sinkhorn) &   29.881 [3.410, 88.351]\\
   LB of Skewness of ICRE (COT; Sinkhorn) &       -0.069 [-0.105, -0.047]\\

   UB of Kurtosis of ICRE (Ours) &  8.690$\times 10^{19}$ [1.926$\times 10^{19}$,2.453$\times 10^{20}$] \\
   LB of Kurtosis of ICRE (Ours) & 0.000 [0.000,0.000] \\

   UB of Kurtosis of ICRE (COT; Sinkhorn)  &   4.807$\times 10^{2}$ [1.893$\times 10$, 1.481$\times 10^{3}$] \\
   LB of Kurtosis of ICRE (COT; Sinkhorn)  & 3.541$\times 10^{-3}$ [8.277$\times 10^{-4}$, 9.519$\times 10^{-3}$] \\
      \hline
    \end{tabular}
    }
\end{table}

\clearpage
\section*{NeurIPS Paper Checklist}

\begin{enumerate}

\item {\bf Claims}
    \item[] Question: Do the main claims made in the abstract and introduction accurately reflect the paper's contributions and scope?
    \item[] Answer: \answerYes{}
    \item[] Justification: Our abstract and introduction accurately reflect the paper's contributions and scope.

\item {\bf Limitations}
    \item[] Question: Does the paper discuss the limitations of the work performed by the authors?
    \item[] Answer: \answerYes{} 
    \item[] Justification: 
    We discuss the limitations of this paper in Section 8.

\item {\bf Theory assumptions and proofs}
    \item[] Question: For each theoretical result, does the paper provide the full set of assumptions and a complete (and correct) proof?
    \item[] Answer: \answerYes{} 
    \item[] Justification: 
    Each theorem, lemma, and corollary has a corresponding assumption.

    \item {\bf Experimental result reproducibility}
    \item[] Question: Does the paper fully disclose all the information needed to reproduce the main experimental results of the paper to the extent that it affects the main claims and/or conclusions of the paper (regardless of whether the code and data are provided or not)?
    \item[] Answer: \answerYes{} 
    \item[] Justification: 
    The paper fully discloses all information necessary to reproduce the main experimental results.

\item {\bf Open access to data and code}
    \item[] Question: Does the paper provide open access to the data and code, with sufficient instructions to faithfully reproduce the main experimental results, as described in supplemental material?
    \item[] Answer: \answerYes{} 
    \item[] Justification: The code is publicly available at \url{https://github.com/NA679723/NeurIPS2026_Joint_Expectation.git}.

\item {\bf Experimental setting/details}
    \item[] Question: Does the paper specify all the training and test details (e.g., data splits, hyperparameters, how they were chosen, type of optimizer) necessary to understand the results?
    \item[] Answer: \answerYes{} 
    \item[] Justification: 
    We provide all experimental details.

\item {\bf Experiment statistical significance}
    \item[] Question: Does the paper report error bars suitably and correctly defined or other appropriate information about the statistical significance of the experiments?
    \item[] Answer: \answerYes{} 
    \item[] Justification: 
    All estimates have confidence intervals.

\item {\bf Experiments compute resources}
    \item[] Question: For each experiment, does the paper provide sufficient information on the computer resources (type of compute workers, memory, time of execution) needed to reproduce the experiments?
    \item[] Answer: \answerYes{} 
    \item[] Justification: 
   All experiments were conducted on an Apple M4 system with 24 GB of memory.
    
\item {\bf Code of ethics}
    \item[] Question: Does the research conducted in the paper conform, in every respect, with the NeurIPS Code of Ethics \url{https://neurips.cc/public/EthicsGuidelines}?
    \item[] Answer: \answerYes{} 
    \item[] Justification: 
    We comply with the NeurIPS Code of Ethics.

\item {\bf Broader impacts}
    \item[] Question: Does the paper discuss both potential positive societal impacts and negative societal impacts of the work performed?
    \item[] Answer: \answerYes{} 
    \item[] Justification:
This work provides theoretical tools for identifying the joint expectation of potential outcomes; misuse risk is mitigated by providing bounds. 
    
\item {\bf Safeguards}
    \item[] Question: Does the paper describe safeguards that have been put in place for responsible release of data or models that have a high risk for misuse (e.g., pre-trained language models, image generators, or scraped datasets)?
    \item[] Answer: \answerNA{} 
    \item[] Justification: The paper poses no such risks.

\item {\bf Licenses for existing assets}
    \item[] Question: Are the creators or original owners of assets (e.g., code, data, models), used in the paper, properly credited and are the license and terms of use explicitly mentioned and properly respected?
    \item[] Answer: \answerYes{} 
    \item[] Justification: 
    We cite the original paper that produced the code package or dataset.

\item {\bf New assets}
    \item[] Question: Are new assets introduced in the paper well documented and is the documentation provided alongside the assets?
    \item[] Answer: \answerNA{} 
    \item[] Justification: 
    The paper does not release new assets.

\item {\bf Crowdsourcing and research with human subjects}
    \item[] Question: For crowdsourcing experiments and research with human subjects, does the paper include the full text of instructions given to participants and screenshots, if applicable, as well as details about compensation (if any)? 
    \item[] Answer: \answerNA{} 
    \item[] Justification: 
    The paper does not involve crowdsourcing nor research with human subjects.

\item {\bf Institutional review board (IRB) approvals or equivalent for research with human subjects}
    \item[] Question: Does the paper describe potential risks incurred by study participants, whether such risks were disclosed to the subjects, and whether Institutional Review Board (IRB) approvals (or an equivalent approval/review based on the requirements of your country or institution) were obtained?
    \item[] Answer: \answerNA{} 
    \item[] Justification: 
    The paper does not involve crowdsourcing nor research with human subjects.

\item {\bf Declaration of LLM usage}
    \item[] Question: Does the paper describe the usage of LLMs if it is an important, original, or non-standard component of the core methods in this research? Note that if the LLM is used only for writing, editing, or formatting purposes and does \emph{not} impact the core methodology, scientific rigor, or originality of the research, declaration is not required.
    \item[] Answer: \answerNA{} 
    \item[] Justification: 
    The core method development in this research does not involve LLMs as any important, original, or non-standard components.

\end{enumerate}

\end{document}